\documentclass[twocolumn,amsmath,amssymb,floatfix,showpacs]{revtex4-1}

\usepackage{graphicx}% Include figure files
\usepackage{dcolumn}% Align table columns on decimal point
\usepackage{bm}% bold math
\usepackage[utf8]{inputenc}
\usepackage{url}
\usepackage{enumitem}
\graphicspath{{figuresFinal/}}

\begin{document}

\title{Network-based model of the growth of termite nests}% Force line breaks with \\

\author{Young-Ho Eom$^{1}$, Andrea Perna$^2$, Santo Fortunato$^{3,4}$, Eric Darrouzet$^5$, Guy Theraulaz$^{6,7}$, Christian Jost$^{6,7,*}$}
\affiliation{{$^1$IMT Institute for Advanced Studies Lucca,
Piazza San Francesco 19, Lucca 55100, Italy} \\
{$^2$Laboratoire Interdisciplinaire des Energies de Demain - Paris Interdisciplinary Energy Research Institute, Paris Diderot University, 10 rue Alice Domon et L\'{e}onie Duquet, Paris, France} \\
{$^3$Complex Systems Unit, Aalto University School of Science, P.O. Box 12200, FI-00076, Finland} \\
{$^4$Center for Complex Networks and Systems Research, School of
Informatics and Computing, Indiana University, Bloomington,
Indiana 47405 USA}
{$^5$IRBI, UMR CNRS 7261, University of Tours, Faculty of Sciences, Parc de Grandmont, 37200 Tours, France} \\
{$^6$Universit\'e de Toulouse, UPS, CRCA (Centre de Recherches sur la Cognition Animale), Bât 4R3, 118 route de Narbonne, F-31062 Toulouse, France} \\
{$^7$CNRS, CRCA,  Bât 4R3 118 route de Narbonne, F-31062 Toulouse, France} \\
{$^*$Corresponding author: CJ (christian.jost@univ-tlse3.fr) }}

\date{\today}% It is always \today, today,

\begin{abstract}
We present a model for the growth of the transportation network
inside nests of the social insect subfamily Termitinae (Isoptera,
termitidae). These nests consist of large chambers (nodes)
connected by tunnels (edges). The model based on the empirical
analysis of the real nest networks combined with pruning (edge
removal, either random or weighted by betweenness centrality) and
a memory effect (preferential growth from the latest added
chambers) successfully predicts emergent nest properties (degree
distribution, size of the largest connected component, average
path lengths, backbone link ratios, and local graph redundancy).
The two pruning alternatives can be associated with different
genuses in the subfamily. A sensitivity analysis on the pruning
and memory parameters indicates that Termitinae networks favor
fast internal transportation over efficient defense strategies
against ant predators. Our results provide an example of how
complex network organization and efficient network properties can
be generated from simple building rules based on local
interactions and contribute to our understanding of the mechanisms
that come into play for the formation of termite networks and of
biological transportation networks in general.
%\\ \\
%Keywords: 3D networks, Termitinae, transportation network, model network, network pruning
\end{abstract}

\pacs{89.75.Hc, 89.75.Fb, 05.65.+b, 87.23.Cc}% PACS, the Physics and Astronomy
\keywords{3D networks, Termitinae, transportation network, model network, network pruning}%Use showkeys class option if keyword

\maketitle

\section{INTRODUCTION}
% transportation: necessary in all biological systems, must be efficient on all scales
Biological transportation networks are a fundamental component of
all living systems, allowing the exchange of information and
material at the scale of the whole system. It is unsurprising that
structures specialized for transportation are found at all levels
of biological organization, including intracellular transportation
along the cytoskeleton, and vascular and neural networks within
the body of animals. At the biological scale of animal groups and
communities, the transportation of food and material, the
movements of animals and the inter-individual encounters are often
supported by specialized transportation networks of trails,
galleries, and burrows that the animals produce and use in their
exploration and foraging movements (reviewed in \cite{perna014}).
Social insects in particular are known to produce some of the most
complex networks of trails and galleries in the animal kingdom,
probably as a result of their high level of sociality. These
networks include trail networks in ants or termites
\cite{buhl009,perna012,griffon015}, and underground systems of
tunnels formed by ants \cite{buhl006} or termites
\cite{lee006,haifig011,jost012}.

% self-organized nature of network growth: from the preferential attachment model (Barabasi, no dimension), only local information, no knowledge of the final product. Such SO models exist for 2D networks (physarum, termite tunnels, trails (deneubourg, perna), road networks (Levinson), ...). No such models for 3D networks (true?) I prefer not to enter into the detail of whether there are 3D networks because there are probably models for angiogenesis (it is a quite large literature) and because all the models based on pruning can work equally well in 3D.
All biological transportation networks share similar functions: by
favoring transportation over distances much larger than those
permitted by simple diffusion they mediate the integration of the
different parts that compose a biological system, supporting the
functional unity of the system as a whole. They also share a
similar morphogenesis, in the sense that almost all biological
transportation networks are produced as the result of
self-organized (SO) morphogenetic processes whereby the growth is
driven by locally available information, in the absence of a
pre-existing master plan of the network
\cite{camazine001,levinson006}.  The formation of animal and human
transportation networks has been modeled with models based on
growth alone (e.g., army ant raid networks \cite{franks91}, ant
gallery networks \cite{buhl006,gautrais014}, or urban street
patterns \cite{strano2012elementary}), pruning of an existing
network \cite{tero006,ma013}, or a combination of both
\cite{valverde009,barthelemy006,barthelemy008}.

% While most studies have developed such SO models in 2D space, some models have also focused on 3D networks \cite{valverde009}.

% network growth in space: models existed from the beginning, to produce networks with similar characteristics, to assess efficiency and sensitivity, ....
A common characteristic of all these networks is that they are
embedded in a 2D or 3D environment, that is, both the position of
network nodes and the layout of network edges are associated with
sets of spatial coordinates. The effects of spatial embedding
cannot be neglected when trying to understand the formation and
topological properties of these networks, because the probability
of existence of a connection between two nodes depends not only on
their relative distance~\cite{Itzkovitz2005}, but also on the
physical arrangement and steric interactions between edges. While
a large part of the existing network literature has dealt with
social or communication networks, which are comparatively less
affected by spatial constraints (see reviews in
\cite{albert002,newman003pub,costa011}), the theoretical
foundations underlying the analysis and the modeling of spatially
explicit networks are comparatively less developed  (see
\cite{barthelemy010pub} for a review) and often deal with specific
fields such as urban transportation and human mobility
patterns~\cite{xie009}.

% Here we develop such a SO-model based on an empirical description of the 3D models of termitinae epiguous nests (chamber-tunnel), transport efficiency necessary for foraging and internal displacement to go to "nice" nest parts.
% Perna's work what is already known (sparse networks, better than random for models that conserve node positions), our model does not conserve node positions, parameterized from local network properties, include two free paras (lambda for peripheral network growth, xsi for pruning according to importance).
% This model will be confronted to the real networks and used to assess the importance of the two parameters on network performance quantified by shortest path, backbone link ratio and local graph redundancy.
In this paper we focus on a specific class of biological
transportation networks represented by the network of chambers and
tunnels that termites of the subfamily Termitinae produce in the
above-ground part of their nest. We identify how structural
features of these transportation networks emerge through
self-organization based on local rules. The nodes of these
networks correspond to chambers and the edges to tunnels
connecting these chambers. These networks provide safe living
space to termites and are connected to an underground tunnel
network (not considered in this study) that connect the nest to
foraging grounds \cite{jmhasly99,noirot00,tschinkel010b}. Termite
movements along these networks involve both bringing back food and
distributing it to the colony (e.g. termite soldiers, larvae and
the royal couple) and daily patterns of movements of various
individuals to different parts of the nest with favorable
environmental conditions.

By analyzing an extended dataset containing nests from three
Termitinae genuses: \emph{Cubitermes}, \emph{Procubitermes} and
\emph{Thoracotermes}, we formulate a model for the growth of the
transportation networks internal to the nests of these different
termites. The model has two free parameters, $\lambda$ that
controls preferential growth on the periphery and $\xi$ that
determines the pruning or removal of existing edges. Two variants
of pruning, random pruning or pruning weighted by betweenness
centrality, are explored, and we compare both variants to random
geometric graphs. We calibrate $\lambda$ and $\xi$ to the real
nests and show that network properties not used in the calibration
process are faithfully reproduced by the models with pruning, the
variant based on betweenness centrality performing slightly
better. The validated models are then used to assess the
sensitivity of some network properties, linked to internal
transportation or nest defense, to the two free parameters in
order to interpret the values that have been found for the real
nests.

%\section{Materials and Methods: from an empirical nest analysis to the nest growth model}
\section{MATERIALS AND METHODS: FROM AN EMPIRICAL NEST ANALYSIS TO
THE NEST GROWTH MODEL}

\subsection{The available termite nests}

The 12 networks modeled in this paper (see Table \ref{tab1})
correspond to the above ground part of termite nests from the
African continent. They all belong to the Termitinae sub-family
that have a common architecture: distinct chambers with a large
diameter interconnected by tunnels of a small diameter (these
different elements can be easily identified because in these nests
the connecting tunnels have a diameter about 10 times smaller than
the chambers' diameter). In our network representation, chamber
barycentres are associated to nodes and tunnels between chambers
to edges between the corresponding nodes. The detailed network
extraction method from x-ray tomographies of the nests has been
described in \cite{perna008a}. Node coordinates are converted from
voxel positions to metric ($x,y,z$) coordinates. The $z$-axis
corresponds to the vertical direction.

The majority of the nests used in our study have been obtained
from natural history museums in France, with their taxonomic
identity only known to the genus level. There are six
\emph{Cubitermes} nest networks (already published in
\cite{perna008a}), four \emph{Procubitermes sjoestedti} networks
collected in 2007 in Côte d'Ivoire, and two \emph{Thoracotermes}
networks (the larger one, a \emph{T. macrothorax}, was collected
in the Republic of the Congo in 2009). See the SM \cite{SM} for
the general shape of all nests. While some aspects of mound
architecture are typical of nests of each genus (mushroom like
shapes in \emph{Cubitermes}, straight pillars in
\emph{Thoracotermes}, and conic forms in \emph{Procubitermes}),
attempts to use this architecture for taxonomy have failed
\cite{noirot70}, pointing to the fact that all the nests in the
Termitinae subfamily (hence all the nests analyzed here) share
similar morphological characteristics and possibly result from
similar construction rules.

% CJ: not yet sure this paragraph is useful, has probably to be rewritten;
% Termitinae are soil feeding termites that enrich the nest construction material with organic matter and exchangeable cations, leading to a high structural stability of these nests \cite{garniersillam95,brauman000,ptacek013} - unless there is a catastrophic event such as a flood torrent or a trampling animal termite mounds do not break down. Termites use these mounds as living space and access through the nest's base an extensive underground network which brings them to their food sources (dead organic matter in or above the soil). Though mound architecture is quite characteristic on the genus level (mushroom like shapes in \emph{Cubitermes}, straight pillars in \emph{Thoracotermes}, conic forms in \emph{Procubitermes}) attempts to use this architecture for taxonomy have failed \cite{noirot70}. Further details on their biology can be found in \cite{grasse84}, a summary for \emph{Cubitermes} can be found in \cite{perna008a,viana013}.
%CJ: I will have to add some more thorough literature review on Procubitermes and Thoracotermes - but rather little is known about biology and ecology of these species.

\begin{table*}[htdp]
\caption{General statistics of the analyzed termite nest networks
and the model parameters that have been calibrated to each nest
(see model description). Nr-C: number of chambers; Nr-T: number of
tunnels; ND: average node degree. LCC: size of the largest
connected component. $(\lambda_{BBP}, \xi_{BBP})$ are the
parameters for the BBP model, $(\lambda_{RP}, \xi_{RP})$ the ones
for the RP model, and $R_{RGG}$ is the parameter for the RGG
model.}
\begin{center}
\begin{tabular}{llccccccccc}
Nest & Genus & Nr-C & Nr-T & ND & LCC & $\lambda_{BBP}$ & $\xi_{BBP}$ & $\lambda_{RP}$ & $\xi_{RP}$ & $R_{RGG}$\\
\hline
C9  &   \emph{Cubitermes}   &   532 &   682 &   2.56    & 507 & 0.021   &   0.117  & 0.020 & 0.126 & 0.199\\
C10 &   \emph{Cubitermes}   &   396 &   371 &   1.87    & 349 & 0.065   &   0.292  & 0.065 & 0.292 & 0.197\\
C11 &   \emph{Cubitermes}   &   344 &   310 &   1.80    & 260 & 0.066   &   0.279  & 0.065 & 0.274 & 0.205\\
C12 &   \emph{Cubitermes}   &   190 &   234 &   2.46    & 183 & 0.067   &   0.246  & 0.066 & 0.244 & 0.274\\
C18 &   \emph{Cubitermes}   &   312 &   343 &   2.2     & 287 & 0.031    &   0.250   & 0.029 & 0.248 & 0.226\\
C19 &   \emph{Cubitermes}   &   295 &   445 &   3.02    & 268 & 0.045   &   0.085   & 0.046& 0.080 & 0.258\\
P67 &   \emph{Procubitermes}    &   1123&   2149&   3.83   & 1091 & 0.025   &   0.0 & 0.025 &0.0 & 0.177\\
P78 &   \emph{Procubitermes}    &   675 &   878 &   2.60   & 598  & 0.014   &   0.046  & 0.014& 0.044 &0.185\\
P79a    &   \emph{Procubitermes}    &   440 &   525 &   2.39  & 347 &   0.006 &   0.290 & 0.006& 0.289 & 0.208\\
P79b    &   \emph{Procubitermes}    &   388 &   383 &   1.97  & 292 &   0.0   &   0.451  & 0.0& 0.447 & 0.203\\
T29 &   \emph{Thoracotermes}    &   98  &   96  &   1.96& 90    &   0.033   &   0.551  & 0.033& 0.515 & 0.303\\
T82 &   \emph{Thoracotermes}    &   1073    &   1470    &   2.74 &
1069 & 0.018 & 0.177 & 0.018 & 0.164 & 0.159 \\ \hline
\multicolumn{6}{r}{mean} & 0.033 & 0.232 & 0.032 & 0.227 & 0.216
\end{tabular}
\end{center}
\label{tab1}
\end{table*}%

\subsection{Empirical network analysis}

\begin{figure}[ht]
\begin{center}
\includegraphics[width=8.4cm]{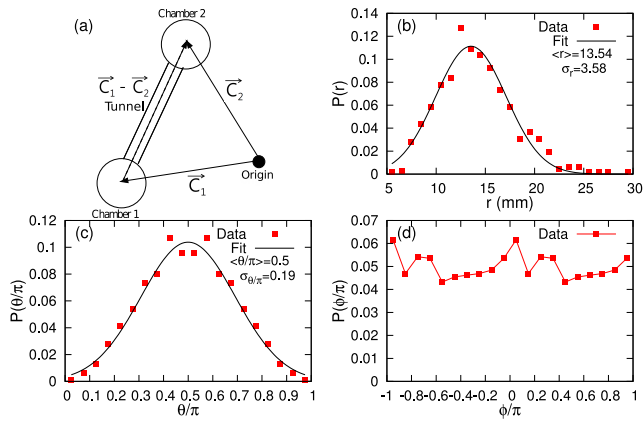}\\
\caption{(Color online) Empirical network description in the case
of nest C9 (\emph{Cubitermes} sp., Central African Republic, see
Fig. 1B in \cite{perna008a}). (a) Each tunnel or edge is
represented as a vector in spherical coordinates, (b) tunnel
lengths ($r$) distribution, (c) distribution of the vertical
component ($\theta$, 0 points upwards and $\pi$ points downwards),
and (d) distribution of the horizontal component ($\phi$). (c) and
(d) are symmetric and periodic respectively because both
$\protect\overrightarrow{c}_{1}-\protect\overrightarrow{c}_{2}$
and
$\protect\overrightarrow{c}_{2}-\protect\overrightarrow{c}_{1}$
are used.} \label{fig:CorridorVec}
\end{center}
\end{figure}

Each tunnel or edge in the nest networks can be considered as a
three dimensional vector represented by Cartesian coordinates
$(x,y,z)$ or by spherical coordinates $(r,\theta,\phi)$. The
latter will be used to characterize the nests. Since there is no
natural orientation in a tunnel connecting chambers
$\overrightarrow{c}_{1}=(x_{1},y_{1},z_{1})$ and
$\overrightarrow{c}_{2}=(x_{2},y_{2},z_{2})$, it can be
represented as either
$\overrightarrow{c}_{1}-\overrightarrow{c}_{2}$ or as
$\overrightarrow{c}_{2}-\overrightarrow{c}_{1}$
(Fig.~\ref{fig:CorridorVec}(a)). Both vectors are used in the
empirical description of the nests, e.g.,
Fig.\ref{fig:CorridorVec}(b-d) for nest C9 that shows the
distributions of tunnel length $r$, the vertical component
$\theta$ and the horizontal component $\phi$. The distributions
for the other 11 nests are shown in the SM \cite{SM}.

Both the $r$ distribution (Fig.~\ref{fig:CorridorVec}(b)) and the
$\theta$ distribution (Fig.~\ref{fig:CorridorVec}(c)) resemble
normal distributions, the latter with mean $\pi/2$. However, the
distribution of $\phi$ ($x-y$ plane) rather follows a uniform
distribution (Fig.~\ref{fig:CorridorVec}(d)). These patterns are
confirmed in the other nests (see SM \cite{SM}). This means that
each termite nest can be characterized by three parameters: the
mean and standard deviation of its tunnel length distribution
$(\bar{r}, \bar{\sigma}_r)$, and the standard deviation of the
vertical $\theta$ direction component $(\bar{\sigma}_\theta)$.

\subsection{\label{sec:level1} Model description}

Based on the above empirical observations we propose a simple nest
growth model, betweenness based pruning (BBP) model, with five
main procedures: (i) set nest boundaries and initial node, (ii)
determine the initiation node for the next tunnel, (iii) construct
an edge (iv) create a new node (and give it an increasing unique
identification number $i \geq 1$) or connect to an existing node
if it is close to the end of the new edge, (v) prune edges of
lesser importance.

\begin{enumerate}[label=(\roman*)]
\item \textbf{Initial and boundary condition:} From the empirical nest data we can compute the x-y-z intervals $[x_{min},x_{max}]$, $[y_{min},y_{max}]$ and $[z_{min},z_{max}]$ for each nest. We impose initial and boundary conditions based on these intervals. We then choose an initial node $(x_{0}, y_{0}, z_{0}) $ randomly in the intervals $x_0 \in [0.75x_{min}+0.25x_{max},0.25x_{min}+0.75x_{max}]$, $y_0 \in [0.75y_{min}+0.25y_{max},0.25y_{min}+0.75y_{max}]$ and $z=z_{min}$. During nest growth we assume boundary conditions which have the shape of an ellipse in the $x-y$ plane. The formula of the ellipse is given by
\begin{equation}
\left(\frac{x-X_c}{A}\right)^2+\left(\frac{y-Y_c}{B}\right)^2=1
\end{equation}

with $X_c=(x_{min}+x_{max})/2$, $Y_c=(y_{min}+y_{max})/2$, $A=\left(x_{max}-x_{min}\right)/2$ and $B=\left(y_{max}-y_{min}\right)/2$. The $z$ coordinate cannot be below $z_{min}$, but there is no upper limit for $z$. The biological rationale behind these choices is that termite nest volumes are proportional to colony size \cite[colony size is the number of individuals living in a single colony]{josens010} and that nests are not enlarged during colony growth but rather rebuilt from scratch - colony size can therefore provide a template for the surface of the initial nest site construction and for the final nest height.

\item \textbf{Determining the node of the next edge's origin:} At
each time step we choose a node randomly from all existing ones.
However, in order to foster nest expansion (growth from peripheral
nodes) we give preference to the latest added nodes by choosing
node $i$ with probability $\exp(-\lambda(n-i))$. Here $n$ is the
total number of nodes for the given nest.
% YHE, please check this formula: the latest added nodes have large i, their probability to be chosen will thus be smaller then the one from the first built nodes with lower i. How did you implement this rule?
The characteristic time constant $\lambda$ is specific to each nest and will be chosen by calibration (see below). The biological idea is that termites mark construction material with some volatile chemical \cite{bruinsma79} and prefer recently built elements to continue construction.

\item \textbf{Direction and length of the next edge:} The length
($r$) and the directions ($\theta$, $\phi$) of the next edge are
chosen randomly from the normal or uniform distributions
determined above. Edge length $r$ is restricted to $r\geq r_{min}$
where $r_{min}$ is the observed minimum edge length (see also the
next rule, shorter edges would lead to an edge connection to the
original node). The vertical component $\theta$ is restricted to
the interval $(0, \pi)$. Furthermore, since nest construction goes
upward and since edges in the original data are not oriented we
avoid excessive downward construction by replacing a $\theta >
\pi/2 + \sigma_\theta$ by $\theta = \pi-\theta$ (recall that
$\theta = \pi$ points downwards). Finally, $\phi$ is drawn
randomly from a uniform distribution in $(0, 2\pi)$.

\item \textbf{Edge construction:} This new edge is only added if
it does not quit the ellipsoid boundary condition (otherwise a new
edge $(r, \theta, \phi)$ is drawn until it remains within these
boundaries). The biological rationale is that termites sense
gravity \cite{jander74} and do not extend construction over empty
space (note that the construction of the characteristic ``hats''
in \emph{Cubitermes} nests is not included in our model). A new
node is created if there is no existing node within the distance
$r_{min}$ of the endpoint of the new edge. If there already exists
a node within this distance the new edge is connected to this
node.

\item \textbf{Pruning of edges:} At each time step $i$ we compute
the edge-betweenness  of each edge~\cite{girvannewman002}, that is
the number of shortest paths between pairs of nodes that pass
through this edge, and remove the one with the smallest value with
probability $\xi$. Such pruning is a common feature of most
observed transportation networks (reviewed in \cite{perna014}). It
has been directly observed in termite's underground tunneling
networks \cite{jost012,bardunias009a}, and there is an indirect
evidence that it happens in \emph{Cubitermes} nest growth
\cite{perna008a}.
\end{enumerate}

Repeat steps (ii-v) until the model has the same number of nodes
as the original nest. The biological rationale underlying this
stop criterion is again the observation that mound size is
proportional to colony size \cite{josens010} and mound size is
better approximated by the number of chambers (volume) than the
number of edges (network length).
% do not use network length, tunnels have very low volume

Since $\lambda$ is related to nest height $h$ and $\xi$ is related to the total number of tunnels $L$, we determine these two parameters one by one, first estimating $\lambda$ (with an arbitrarily fixed $\xi$) by minimizing the error function
\begin{equation}
\epsilon_1 = |h_{data}-h_{model}|/h_{data} \label{eq:error1}
\end{equation}
with a standard bisection method,  then estimating $\xi$ (fixing the already estimated $\lambda$) by minimizing the error function
\begin{equation}
\epsilon_2 = |L_{data}-L_{model}|/L_{data}.
\end{equation}
For each value of $(\lambda, \xi)$ we simulate 500 nests in order to compute mean $h_{model}$ and $L_{model}$. Overall error is defined as $\epsilon = \epsilon_1 + \epsilon_2$. We will use an analysis of variance to test whether the estimated parameters are specific for a genus, reporting the resulting $F$ statistic with the associated degrees of freedom and $p$-values.

\subsection{Alternative models: Random geometric graph (RGG) model and random pruning (RP) model}
We consider two alternative models to clarify the performance of
the BBP model and the roles of local rules in the BBP model: (i)
random geometric graph model and (ii) random pruning model. A
random geometric graph (RGG) is the mathematically simplest
spatial network~\cite{barthelemy010pub}. The RGG model for a given
nest is embedded in the cylindrical space having the same interval
of $[x_{min}, x_{max}]$, $[y_{min},y_{max}]$, and $[z_{min},
z_{max}]$ obtained from the nest with the ellipsoid boundary
condition given by Eq. (1). The nodes of the RGG model are
uniformly distributed in this cylinder. Two nodes $i$ and $j$ are
connected if the below condition is satisfied:
\begin{equation}
\left(\frac{X_i-X_j}{A}\right)^2+\left(\frac{Y_i-Y_j}{B}\right)^2+
\left(\frac{Z_i-Z_j}{C}\right)^2\leq R_{RGG}^2
\end{equation}

Here $X_i$, $Y_i$, and $Z_i$ indicate x, y, and z coordinates of
node i, respectively and $A=(x_{max}-x_{min})/2$,
$B=(y_{max}-y_{min})/2$, and $C=(z_{max}-z_{min})/2$. We determine
the value of $R_{RGG}$ such that it gives us the same number of
links in the RGG network as in the original nest network.

The random pruning (RP) model is the same as the original model in
every points (i.e., from (i) to (iv) in the previous subsection)
except the pruning process (i.e, (v) in the previous subsection)
where we remove a randomly selected link with probability $\xi$
rather than the lowest edge-betweenness link.

\subsection{Model validation}

We will use five emergent properties to compare between the
original and the simulated networks: (i) node degree distribution,
(ii) size of the largest connected component, (iii) average
topological path length in the largest connected component, (iv)
backbone link ratio (fraction of edges whose removal leads to a
disconnection of the largest connected component), and (v) local
graph redundancy (as defined in \cite{perna008b}, it complements
backbone link ratio by computing the mean of the inverse of the
topological path length to connect two adjacent nodes once the
direct link has been blocked: low values indicate long detours).
Predicted properties will be based on 1000 simulated networks.
%(YHE: how
%could we quantify the difference between observed and predicted
%networks? There are statistical tools that could be applied nest
%by nest - Kolmogorov-Smirnoff to compare between distributions,
%one-sample t-tests for average path length, total network length
%and backbone link ratio - but I am afraid this would make
%difficult reading? To be discussed).
%
%(\textbf{Answer: I agree with you. Since the points of paper are
%to provide (i) statistical analysis of link construction and (ii)
%a simple self-organized network model in three dimensional space
%showing qualitatively similar behaviors with empirical nests, the
%model doesn't need to fit perfectly with empirical nests.})

\subsection{Sensitivity analysis}
% define the performance parameters: Average distance, backbone link ratio, local graph redundancy.
% describe how you varied lambda, xi and N (number of chambers) to explore how these performance parameters depend on lambda and xi
We assess the influence of $\lambda$, $\xi$, and network size
(number of nodes) on four network properties: (i) size of the
largest connected component, (ii) average topological distance
between any two nodes of the largest connected network component,
(iii) backbone link ratio, and (iv) local graph redundancy. For an
efficient transport inside the nest and easy defense against
predators (such as ants) termite nests should show low average
distance (fast transport), a high backbone link ratio (tree-like
structures, tunnel blocking by a soldier efficiently isolates a
part of the nest), and a low local graph redundancy (blocking of a
tunnel forces attacking ants to take long detours). The values
chosen for $\lambda$ and $\xi$ cover the range of the estimated
values (from 0 to 0.4 in steps of 0.01), while the number of nodes
cover the original nest sizes (200, 400, 600 and 1000 nodes).

\section{RESULTS}
% structure of Results section
% (a) general nest fit (eps) and model validation
% (b) estimation of lambda/xsi and their dependence on genus,
% (c) sensitivity analysis

\subsection{Model validation}

\begin{figure*}[hbt!]
\begin{center}
\includegraphics[angle=-90,width=\textwidth]{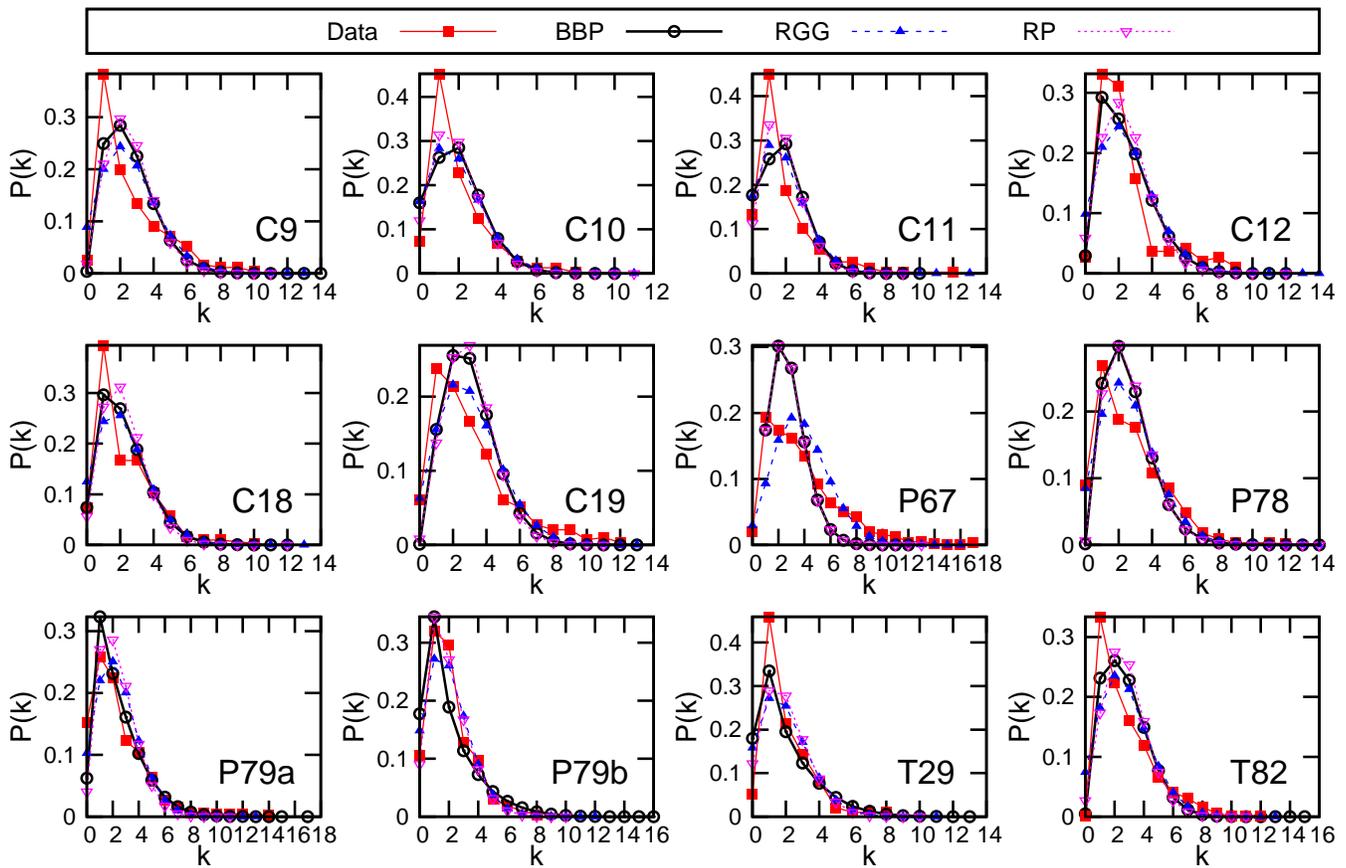}
\caption{(Color online) Degree distribution of the real networks
(Data: red line) and the networks generated by the BBP model (BBP:
black line), by the RGG model (RGG: blue line), and RP model (RP:
magenta line). $P(k)$ are relative frequencies (summing to 1). }
\label{fig:degDist}
\end{center}
\end{figure*}

\begin{figure}[hbt!]
\begin{center}
\includegraphics[width=38mm]{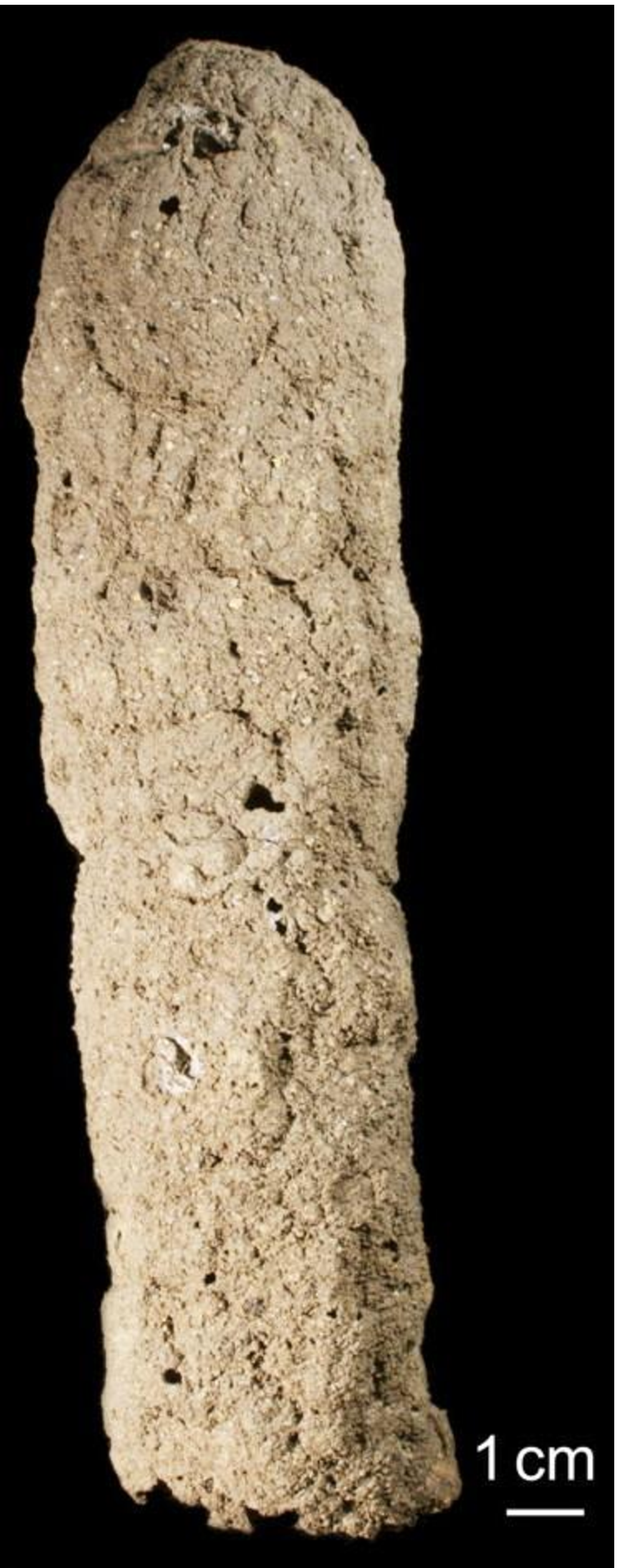}
\includegraphics[angle=0,width=22mm]{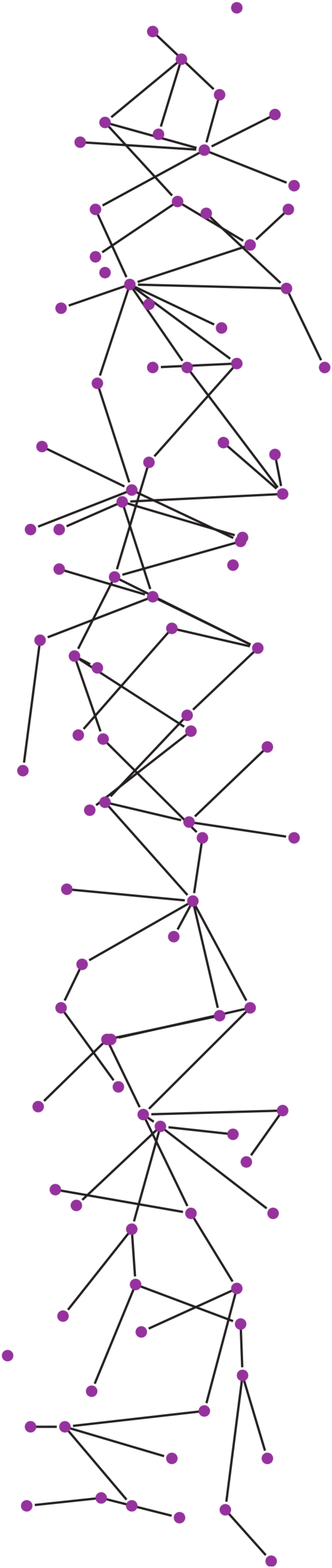}
\includegraphics[angle=0,width=23mm]{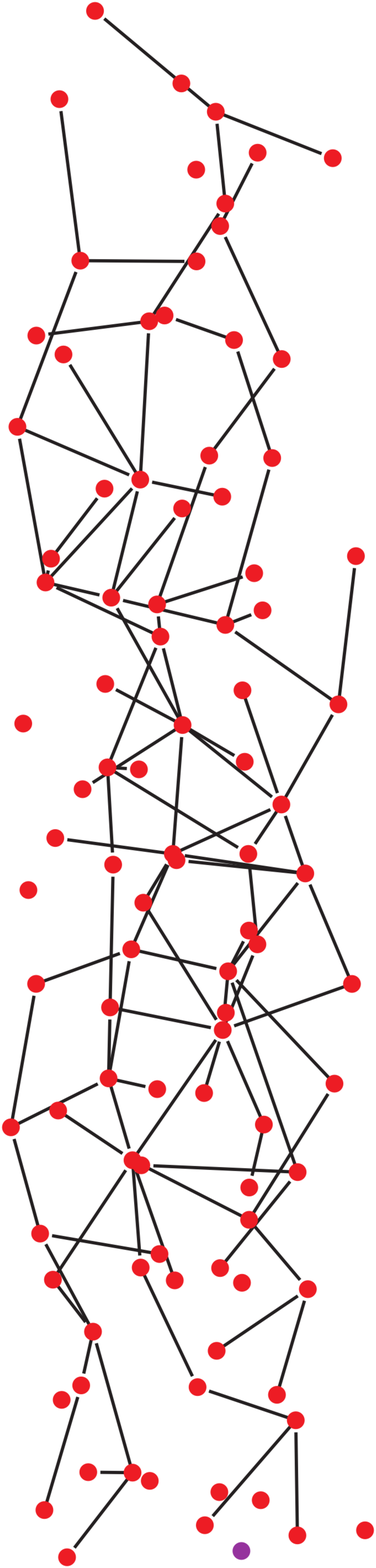}
\caption{(Color online) Analyzed \emph{Thoracotermes} nest T29
with its network representation (middle) and a simulated network
(right) by the BBP model. The pink node (lower right) indicates
the initial node of the simulated network. Note that while there
are isolated chambers in the original network, isolated chambers
are more prevalent in the simulated nests. See the Discussion for
further comments.} \label{fig:t29model}
\end{center}
\end{figure}

Figure~\ref{fig:degDist} shows the degree distributions of the
real networks and the simulated networks generated by the BBP
model, RGG model, and RP model for each nest. The real networks
and the simulated networks show qualitatively similar behaviors,
with peaks around $k=1,2$ and exponential type decay of the right
tail. However, all of the nests have a peak at $k=1$ while the BBP
model creates a peak at $k=2$ for seven nests: the BBP model seems
to create less dead ends than there exist in the real networks
(see also Fig.~\ref{fig:t29model}). Nest P67 shows the worst fit
with a larger tail in the model networks: note that this nest
looks like two nest parts that have fused together during growth
(see SM Fig.~6), a process not considered in the simulated
networks. The degree distributions generated from the RGG model
and the RP model show similar patterns as the distributions from
the BBP model.

\begin{figure*}[hbt!]
\begin{center}
\includegraphics[angle=-90,width=\textwidth]{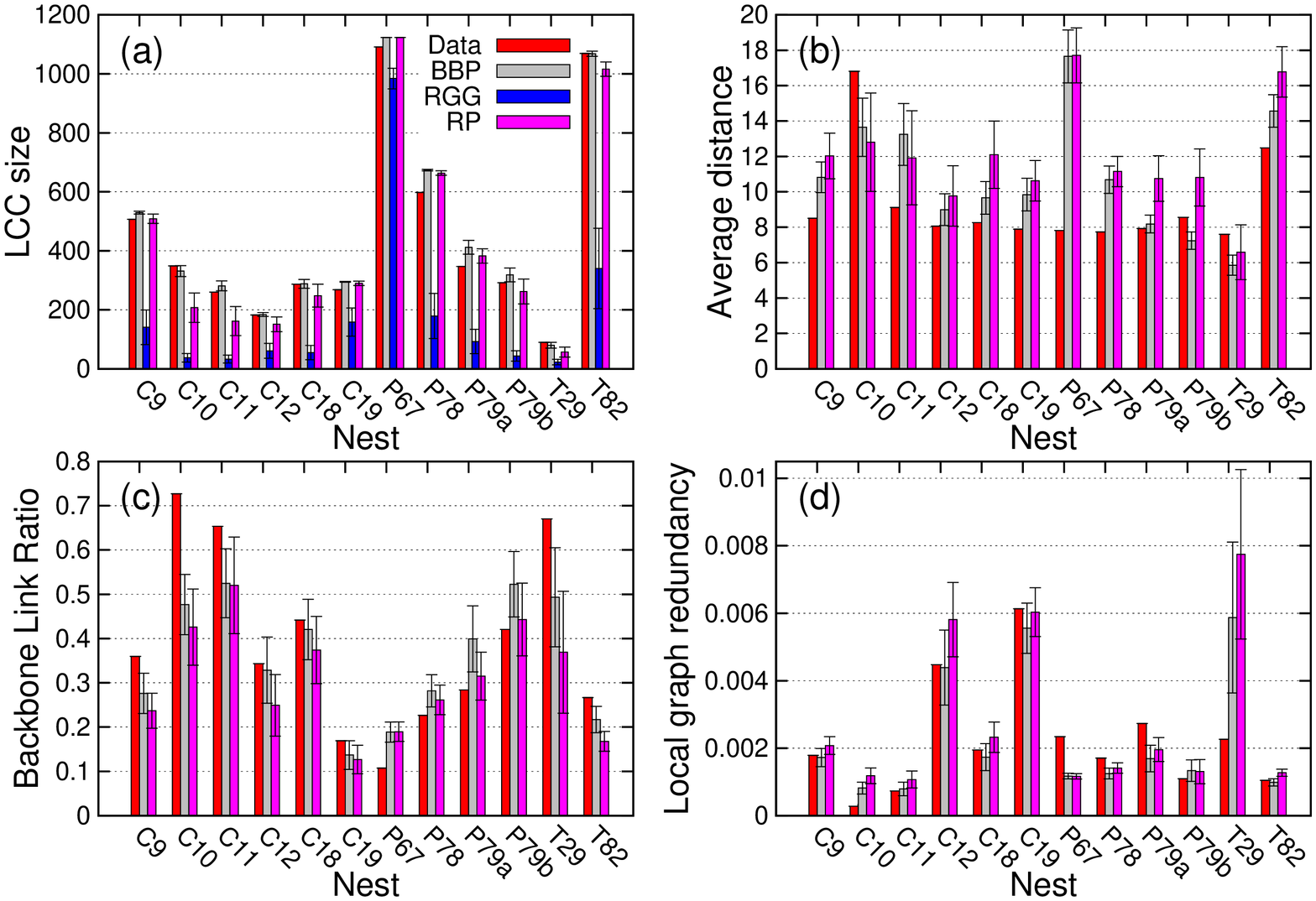}
\caption{(Color online) Comparison between the real networks
(Data: red bar) and the networks generated by the BBP model (BBP:
grey bar), by the RGG model (RGG: blue bar), and RP model (RP:
magenta bar). (a) Size of the largest connected components (LCC).
(b) Average topological distance between any two nodes in the LCC.
(c) The backbone link ratio. (d) Local graph redundancy. The error
bars represent standard deviations computed from 1000 model
generated networks.} \label{fig:validation}
\end{center}
\end{figure*}

The largest connected component (LCC) of the network is
fundamental for internal transportation since communication is not
possible between disconnected components.
Figure~\ref{fig:validation}(a) represents the sizes of the largest
connected components in real networks and simulated networks. The
BBP model generated networks having comparable sizes of the
largest connected components with the ones in real nests for most
cases while the RGG model failed to generate the largest connected
components of proper size in most cases. However, the performance
of the RP model is comparable with the BBP model. It is notable
that the discrepancy between real networks and networks generated
by RP models is larger when the average degree is low. The blind
cutting of random pruning can in this case increasingly affect
important links and thus reduce LCC, while with higher average
degrees many 'back-up' links exist that help preserve the LCC.
% CJ: I reformulated your phrase
%Because, when average degree is low, random pruning blindly cuts
%important links keeping the largest connected components while
%average degree is high, since there are many links 'back-up' links
%to keep the largest connected component, the largest connect
%component is preserved.
We can observe this pattern also in the sensitivity analysis of
the RP model (see Fig. 13 in \cite{SM}): with increasing
probability to prune an edge ($\xi$) LCC quickly degrades, while
it is better preserved in the BBP model (see
Fig.~\ref{fig:sensAna} below).
% CJ: I added a phrase how this can be observed in the sensitivity analysis
This indicates that betweenness based pruning is a better strategy
than random pruning to keep the largest connected component.

Figure~\ref{fig:validation}(b) compares the average topological
distance between any two nodes in the largest connected component,
Fig.~\ref{fig:validation}(c) the backbone link ratios, and
Fig.~\ref{fig:validation}(d) the local graph redundancy.
% CJ: added the next phrase
The RGG statistics are not shown because computing them makes
little sense when the LCC is too small. Overall both the BBP and
RP model successfully reproduced the average distances, backbone
link ratios, and local graph redundancies of the real networks.
% CJ: you cannot talk here of RGG, its not plotted in these figures
% while the performance of the
% RGG and RP model are relatively poorer than the BBP model.
% CJ: I attenuate the previous phrase with the correlation information
However, in the case of average distance neither BBP nor RP catch
the variation between nests: Kendall's correlation coefficient
$\tau$ with the real nests ($\tau_{BBP}=0.30$ and
$\tau_{RP}=0.363$) is not significantly different from 0. In the
case of LCC ($\tau_{BBP}=0.939$ and $\tau_{RP}=0.818$), backbone
link ratio ($\tau=0.606$ for both models) or local graph
redundancy ($\tau_{BBP}=0.636$ and $\tau_{RP}=0.606$) $\tau$ is
significantly different from 0.
% CJ: Young-Ho, can you compute Kendall's correlation coefficients for local graph redundancy and add the values? thanks.
Given the plotted standard deviations notable differences can only
be detected for nests C9, P67, P78, and T29 concerning average
distance, nests C10 and P67 concerning backbone link ratio, and
C10 and P67 concerning local graph redundancy. Again, nest P67
stands out in this comparison. See \cite{SM} for the full
distributions of distances as in Fig.~\ref{fig:degDist}.

To compare the performances of the BBP model and the RP model
quantitatively we define a $Z$-score for each metric $X$ such
that:
\begin{equation}
Z(X)= \frac{X_{real}-\langle X\rangle_{Model}}{\sigma_{Model}}
\end{equation}
where $\langle X \rangle_{Model}$ is the average of X for the
given model and $\sigma_{Model}$ is the standard deviation. We
show the Z-scores for each metric and each nest in Tables I-IV
in~\cite{SM}. Interestingly we found that the BBP model has a
tendency to perform better than the RP model for \emph{Cubitermes}
and \emph{Thoracotermes} nests while for the nests of
\emph{Procubitermes} the RP model performs better than the BBP
model.

\subsection{Calibrated parameters}

For each nest (except P67 and P79b) we found parameter sets
$(\lambda_{BBP}, \xi_{BBP})$ with $\epsilon<0.025$ for the BBP
models and $(\lambda_{RP}, \xi_{RP})$ with $\epsilon<0.027$ for
the RP model.
% CJ: Young-Ho, I do not know the epsilon values for the RP fits - can we generalize this phrase for both models?
Nest P67 with $\epsilon_{BBP}=0.285$ and $\epsilon_{RP}=0.283$ is
again an exception. In the case of Nest P79b, the BBP model (
$\epsilon_{BBP}=0.125$) and the RP model ($\epsilon_{RP}=0.129$)
provide us the taller nests than real ones since the errors
$\epsilon$ are mainly from the height of the models (i.e., from
$\epsilon_1$ in Eq.~\ref{eq:error1}). Table \ref{tab1} summarizes
the principal nest properties and the estimated parameter sets.

Model parameter $\xi$ does not depend on taxonomy (analysis of
variance at the genus level, $F=0.78, df=(2,9), p=0.49$ for BBP
and $F=0.60, df=(2,9), p=0.57$ for RP), but $\lambda$ depends on
Genus ($F=6.61, df=(2,9), p=0.017$ for BBP and $F=6.16, df=(2,9),
p=0.021$ for RP; Tukey post-hoc: \emph{Cubitermes} has
significantly higher $\lambda$ than \emph{Procubitermes} and
\emph{Thoracotermes}).

% (c) Sensitivity analysis, description of general tendencies (average distance does not depend on xsi, backbone link ratio and local graph redundancy can be optimized together, placement of the original nests in these figures.

\subsection{Sensitivity to parameters $(\lambda,\xi)$}

\begin{figure*}[p]
\begin{center}
\includegraphics[width=\textwidth]{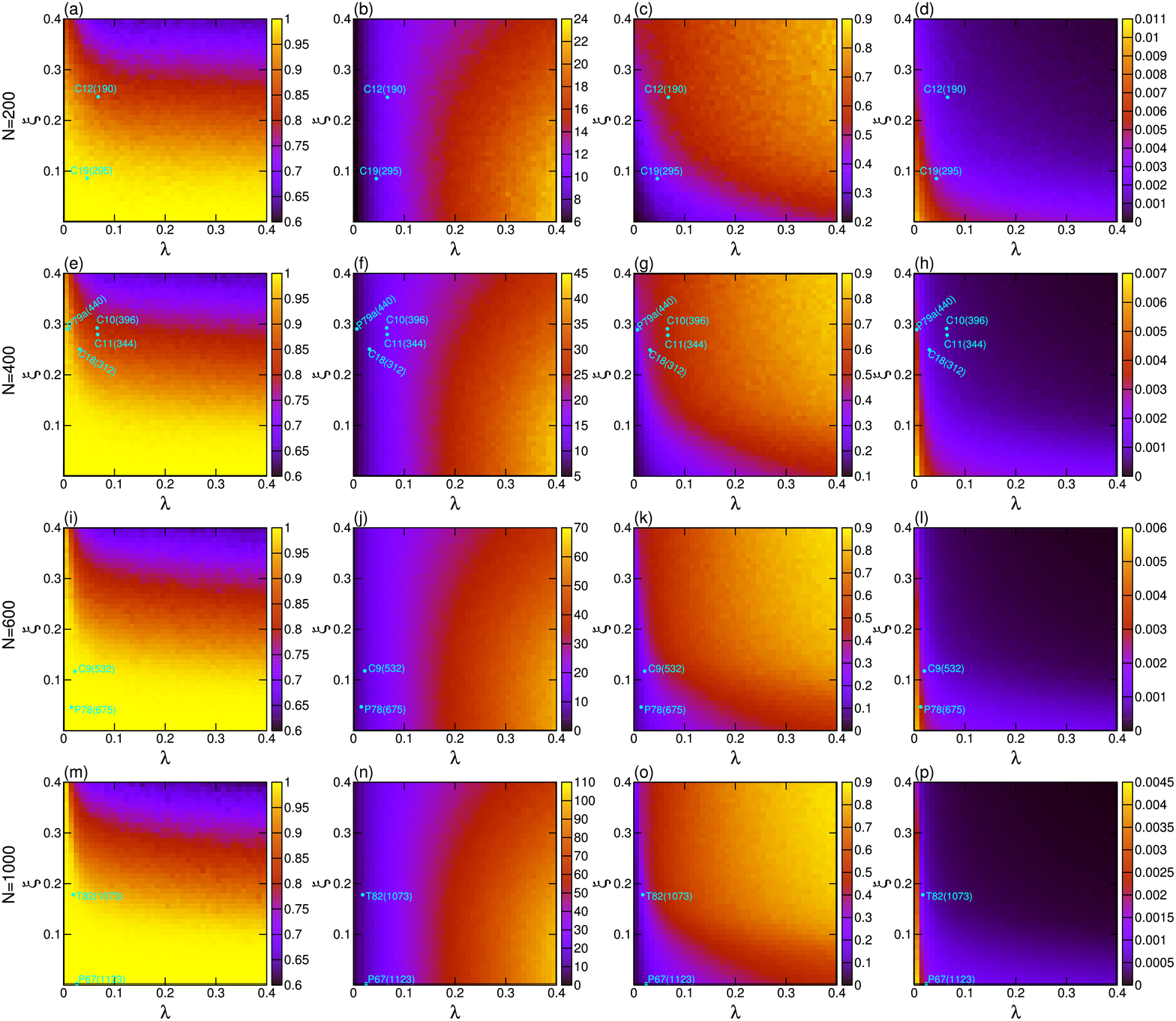}
\caption{(Color online) Sensitivity analysis of the BBP model. The
sizes of largest connected components (LCC) are depicted in (a),
(e), (i), and (m). LCC is shown as the fraction of the original
network size. The average distances are depicted in (b), (f), (j),
and (n). The backbone link ratios are depicted in (c),(g),(k), and
(o). The local graph redundancies are depicted in (d), (h), (l),
and (p). Note that nest P79b with $\xi = 0.451$ and nest T29 with
$\xi=0.551$ lie outside the range of simulated $\xi$-values, they
therefore do not appear in figures (e) to (h) and (a) to (d)
respectively. The spatial information of Nest C9 was considered as
physical constraints for the analysis.} \label{fig:sensAna}
\end{center}
\end{figure*}

Figure~\ref{fig:sensAna} shows the sensitivity of network
properties to the two free parameters $(\lambda_{BBP},
\xi_{BBP})$. We can observe that for the given parameter space,
the size of the largest connected component is preserved when
$\xi<0.3$. We further see that average distance only depends on
the decay rate $\lambda$, with low distances for low values of
$\lambda$. Note that both sensitivities are quite different in the
BP model (Fig. 13 in \cite{SM}): the network quickly breaks down
with increasing $\xi$, especially for large $\lambda$, leading to
small LCC's and low average distances in these LCC's. The nests
have actually all values of $\lambda < 0.1$, indicating a moderate
preference to continue construction from the most recent nodes.
Backbone link ratio and local graph redundancy have a more complex
dependence on $(\lambda, \xi)$ (isolines seem to be linked to the
product $\lambda \xi$), but termites could obtain a high backbone
link ratio and a low local graph redundancy by increasing both
$\lambda$ and $\xi$. All these qualitative observations are
independent of nest size (number of nodes).
% maybe to be moved to Discussion
The fact that the analyzed nests have nevertheless $\lambda < 0.1$
indicates that low average distances have more importance than
increasing backbone link ratio or decreasing local graph
redundancy. Only increasing $\xi$ would further optimize the last
three criteria, but also produce more disconnected chambers
(decrease LCC) which incurs a cost to the colony (construct living
space that cannot be used).

\section{DISCUSSION}

% refs for pruning: perna008b, latty011, jost012, bardunias009a

% Suggestions for "fil conducteur" in the discussion\\
% a - recall the major goal of the paper: suggest a SO network growth model based on the empirical nest analyses.\\
In this paper we develop a simple network growth model (the BBP
model) to test how the nest architecture of termites in the
Termitinae sub-family emerges from self-organization based on
local rules only. These nests consist of spherical chambers
connected by tunnels, an architecture that can be represented as a
network. The BBP model uses empirical nest descriptions (edge
length and orientation) and two free parameters that control
peripheral growth and pruning of existing edges. The free
parameters are calibrated by fitting the model to the observed
nest height and the number of edges.

% b - overall this has worked (summarize network properties that are well reproduced), important are correct combination of peripheral growth ($\lambda$) and pruning ($\xi$) - discuss relevance and alternatives for our optimization criterion $\epsilon$. Discussion of these two features in other spatial networks in the literature.\\
The BBP model correctly reproduces several emergent properties: a)
node degree distribution (Fig.~\ref{fig:degDist}), b) the size of
the largest connected component (Fig.~\ref{fig:validation}(a)), c)
the average distance between any two nodes in the largest
connected component (Fig.~\ref{fig:validation}(b)), d) the
backbone link ratio (Fig.~\ref{fig:validation}(c)), and e) the
local graph redundancy (Fig.~\ref{fig:validation}(d)). The
properties (c) to (e) are of ecological relevance: the populations
move around in these nests to place brood or larvae in chambers
with optimal climatic conditions or to store and retrieve food
(dead organic matter), short distances are therefore useful. It
has already been shown for \emph{Cubitermes} nests that their
average distances are shorter than what could be obtained by
randomly connecting the existing nodes \cite{perna008b}, and our
model correctly reproduces these lengths for most analyzed nests.
The nest must also protect the colony against ant predators: this
is done by the soldiers who can block a tunnel with their head
capsule. A treelike network structure (or high backbone link ratio
combined with local graph redundancy) helps with this strategy,
and our model also correctly predicts these properties.

% comparison between BBP and RP/RGG
We compare our BBP model to two alternatives, the RP model that is
identical to BBP but with pruning applied to randomly chosen edges
instead of the smallest betweenness centrality edge as in BBP, and
to the Random Geometric Graph (RGG) model with the same space
constraints as the two previous models. The latter can be ruled
out because it cannot reproduce the observed largest connected
component (LCC). The RP model generally performs as well as the
BBP model (in the case of \emph{Procubitermes} even better
according to the Z-scores) for nests with average degree $\geq
2.0$. However, in nests with lower average degree
(C10,C11,P79b,T29) RP tends to give fragmented networks.
Furthermore, the sensitivity analysis shows that LCC size is
conserved in the BBP model for a much larger range of pruning
probabilities than in the RP model. Overall, pruning based on
betweenness centrality (BBP) reproduces networks similar to the
real networks more robustly than the RP model, we therefore
conclude that betweenness centrality based pruning might be an
important mechanism in termite nest construction.

% \textbf{Andrea, can you think of a paragraph placing these properties, peripheral growth and pruning, in a wider literature context?}.
While our model is intended to specifically reproduce the general
properties and appearance of termite nests of the Termitinae
sub-family, we can speculate that the key ingredients that
regulate network morphogenesis in our model (peripheral growth and
pruning) are likely to be shared also by a large number of
biological transportation networks. For instance, pruning
phenomena are observed in the maturation of neural networks (e.g.
through programmed cell death \cite{oppenheim1991cell} and
synaptic pruning \cite{Chechik1998}). Mycelial networks formed by
fungi similarly undergo a maturation process that involves
peripheral growth of the hyphal tips in response to local changes
of turgor pressure~\cite{Lew2011} and regression of filaments from
nutrient depleted regions~\cite{falconer2005biomass}. Growth of
peripheral filaments that are subsequently pruned is also central
to the formation of the network of cytoplasmic tubes that
constitute the body of the plasmodium of \textit{Physarum
polycephalum}~\cite{kessler82}. It seems plausible that
self-organized network construction and optimization requires
these mechanisms to operate on an initially highly connected
network that is subsequently pruned. This could respond for
instance to the fact that self-organized mechanisms can only
evaluate transportation performance through the transportation
itself, and the fine-tuning and optimization of the network
requires to operate on already formed connections.

The sensitivity analysis (Fig.~\ref{fig:sensAna}) indicates that
our two free parameters $\lambda$ (for peripheral growth) and
$\xi$ (for pruning) strongly influence both average distance and
backbone link ratio (as well as local graph redundancy, another
measure of how efficiently the nest can be defended, see
\cite{perna008b}). We calibrated them by concentrating on a
mini\-mal number of features, nest height and total number of
tunnels (i.e., links), in order to use the other nest features as
emergent properties for model validation. The good qualitative
agreement between these emergent features and the original nest
features indicates that the estimated parameter values  would not
change much if calibration had been based on more features. The
peripheral growth parameter $\lambda_{BBP}$ has mean value 0.033,
meaning that the probability for a given node to be chosen for the
next edge is divided by half every 30 newly added chambers.
Typical \emph{Cubitermes} nests grow in 1-3 months
\cite{noirot70}, meaning that the half life is of the order of
some days to a week, a plausible duration for chemical marking in
termites. In larger nests our model will also predict that growth
becomes spotted on the nest surface, in agreement with field
observations by one of the authors (CJ, unpublished).

% c - its an empirical model, not an IBM, so the proximate mechanisms how termites build their nest remain to be studied. Discuss other relevant empirical network models in social insects (ants, termites). \\
Our model does not explain how termites decide to dig a new tunnel
or how they choose an orientation in space - it is an empirical
model at an intermediate scale. As such it resembles the 2D ant
tunneling network models suggested by
\cite{buhl006,toffin009,gautrais014}, or 2D termite tunneling
models as suggested by \cite{lee006,haifig011,jost012}. However,
though 3D network data of social insect nests become increasingly
available \cite{tschinkel010,tschinkel011,minter012}, our model
seems to be the first to predict the nest's 3D network
architecture. Further work should investigate how nests grow in
time \cite{minter012} and, on the other end, how termites decide
where and when to construct.

Another feature not predicted by our model are the characteristic ``hats'' atop the \emph{Cubitermes} nests (see SM \cite{SM}). It is not known how termites decide to start constructing laterally, and we could not identify statistical properties specific to the height where the hat is constructed. The behavioral algorithm underlying this specific ``hat'' feature is therefore an open question.

% d - some words about what the model does not explain (connection to underground network (rather 2D) to avoid disconnection of lower nodes, too many dead ends, hats in Cubitermes, multi-column nests such as P67)\\
There were also some misfits in this work. For example, nest P67
is badly explained. This nest actually consists of two columns
that have grown together (see SM \cite{SM}). This bad fit might be
corrected by letting the model grow from two randomly chosen
initial nodes. Also, the above ground nest is only part of a
termite colony, it connects to an extensive underground tunneling
network through which termites access food. The absence of this
underground network in our data leads to an underestimation of
edge-betweenness of the lower edges, thus explaining the often
observed pruning of these edges (Fig.~\ref{fig:t29model}) that
leads to an illogical disconnection between the above and below
ground nest parts. This could only be corrected if one collects,
in addition to the above ground part, a cast of the corresponding
underground network~\cite{tschinkel011}. Note also that
underground tunnels are often built at a constant distance below
ground, thus effectively leading to a 2D below ground network
\cite{jmhasly99,moreira004}: an extended dataset could therefore
explore how 3D networks connect to 2D networks and how this alters
network properties. A further discrepancy is observed in the node
degree distributions (Fig.~\ref{fig:degDist}): all nests have a
peak at degree 1, but for half the nests the model predicts a peak
at degree 2. This could probably be corrected by including this
criterion when fitting $\lambda$, or by treating the fusion
distance $R$ (model part iv) as a free parameter. In the interest
of keeping the model simple we refrained from such extended
fitting procedures.

% e - sensitivity analysis: termites show weak tendency for peripheral growth only ($\lambda$ small) and some upper limit for pruning (though $\xi$ can become very high) probably to avoid too many isolated chambers. Discuss relevant "optimality" features in other networks in the literature. \\
The sensitivity analysis (Fig.~\ref{fig:sensAna}) shows that
average distance only depends on the peripheral growth parameter
$\lambda$, not on the pruning parameter $\xi$. Backbone link ratio
and local graph redundancy, on the other side, depend on both
parameters (higher $\lambda$ can be compensated by lower $\xi$ or
the other way around). Both observations are true for all analyzed
nest sizes. An efficient termite nest should have small average
distances, high backbone link ratios and low local graph
redundancy. The detected nest positions suggest that short
distances are more important for the colony than the other two
criteria (that can be linked to nest defense). However, further
information is required on what termites actually do in their
nests and on their vulnerability to predation before further
speculating about such issues.

% f - "Take-home message": we found a reasonable and parsimonious empirical SO network growth model, peripheral growth and pruning are important for these spatial networks, termites seem to favor short communication distances and less preparation for predator attacks.
In sum, we found a parsimonious empirical network growth model based on self-organized principles that successfully predicts the nest architecture of Termitinae nests. Peripheral growth (i.e. some volatile chemical marking of new chambers/nodes/edges) and pruning of less important edges are important ingredients in this model.

%\textbf{Acknowledgments:}
\section*{ACKNOWLEDGMENTS}
Y.-H. Eom acknowledges support of the EC FET project SIMPOL (No.
610704) and MULTIPLEX (No. 317532). We thank A Robert (IRD, Bondy)
for \emph{Thoracotermes} nest T82, A Nel (National Museum of
Natural History, Paris) for some \emph{Cubitermes} nests  and nest
T29, and P Annoyer (SANGHA) for the other \emph{Cubitermes} nests.
We also thank the hospitals of Toulouse (Pr Rousseau) and Tours
(Pr D Herbreteau) for scanning our nests. Finally, the authors
thank Dr A. Akpesse for helping to collect the
\emph{Procubitermes} nests in Côte d'Ivoire. This work was
supported by the French ANR grant ANR-06-BYOS-0008.

\bibliographystyle{apsrev4-1}
\bibliography{YHE-citations}

%Merlin.mbs v4.21 2009-07-09.
\begin{thebibliography}{10}%
\makeatletter
\providecommand \@ifxundefined [1]{%
 \ifx #1\undefined \expandafter \@firstoftwo
 \else \expandafter \@secondoftwo
\fi
}%
\providecommand \@ifnum [1]{%
 \ifnum #1\expandafter \@firstoftwo
 \else \expandafter \@secondoftwo
\fi
}%
\providecommand \enquote [1]{``#1''}%
\providecommand \bibnamefont  [1]{#1}%
\providecommand \bibfnamefont [1]{#1}%
\providecommand \citenamefont [1]{#1}%
\providecommand\href[0]{\@sanitize\@href}%
\providecommand\@href[1]{\endgroup\@@startlink{#1}\endgroup\@@href}%
\providecommand\@@href[1]{#1\@@endlink}%
\providecommand \@sanitize [0]{\begingroup\catcode`\&12\catcode`\#12\relax}%
\@ifxundefined \pdfoutput {\@firstoftwo}{%
 \@ifnum{\z@=\pdfoutput}{\@firstoftwo}{\@secondoftwo}%
}{%
 \providecommand\@@startlink[1]{\leavevmode\special{html:<a href="#1">}}%
 \providecommand\@@endlink[0]{\special{html:</a>}}%
}{%
 \providecommand\@@startlink[1]{%
  \leavevmode
  \pdfstartlink
   attr{/Border[0 0 1 ]/H/I/C[0 1 1]}%
   user{/Subtype/Link/A<</Type/Action/S/URI/URI(#1)>>}%
  \relax
 }%
 \providecommand\@@endlink[0]{\pdfendlink}%
}%
\providecommand \url  [0]{\begingroup\@sanitize \@url }%
\providecommand \@url [1]{\endgroup\@href {#1}{\urlprefix}}%
\providecommand \urlprefix [0]{URL }%
\providecommand \Eprint[0]{\href }%
\@ifxundefined \urlstyle {%
  \providecommand \doi [1]{doi:\discretionary{}{}{}#1}%
}{%
  \providecommand \doi [0]{doi:\discretionary{}{}{}\begingroup
  \urlstyle{rm}\Url }%
}%
\providecommand \doibase [0]{http://dx.doi.org/}%
\providecommand \Doi[1]{\href{\doibase#1}}%
\providecommand \bibAnnote [3]{%
  \BibitemShut{#1}%
  \begin{quotation}\noindent
    \textsc{Key:}\ #2\\\textsc{Annotation:}\ #3%
  \end{quotation}%
}%
\providecommand \bibAnnoteFile [2]{%
  \IfFileExists{#2}{\bibAnnote {#1} {#2} {\input{#2}}}{}%
}%
\providecommand \typeout [0]{\immediate \write \m@ne }%
\providecommand \selectlanguage [0]{\@gobble}%
\providecommand \bibinfo [0]{\@secondoftwo}%
\providecommand \bibfield [0]{\@secondoftwo}%
\providecommand \translation [1]{[#1]}%
\providecommand \BibitemOpen[0]{}%
\providecommand \bibitemStop [0]{}%
\providecommand \bibitemNoStop [0]{.\EOS\space}%
\providecommand \EOS [0]{\spacefactor3000\relax}%
\providecommand \BibitemShut [1]{\csname bibitem#1\endcsname}%
%</preamble>
\bibitem{perna014}%
  \BibitemOpen
  \bibfield{author}{%
  \bibinfo {author} {\bibfnamefont{A.}~\bibnamefont{Perna}}\ and\ \bibinfo
  {author} {\bibfnamefont{T.}~\bibnamefont{Latty}},\ }%
  \bibfield{journal}{%
  \Doi{10.1098/rsif.2014.0334}{\bibinfo {journal} {Journal of the Royal Society
  Interface}}\ }%
  \textbf{\bibinfo {volume} {11}},\ \bibinfo {pages} {20140334} (\bibinfo
  {year} {2014})%
  \bibAnnoteFile{NoStop}{perna014}%
\bibitem{buhl009}%
  \BibitemOpen
  \bibfield{author}{%
  \bibinfo {author} {\bibfnamefont{J.}~\bibnamefont{Buhl}}, \bibinfo {author}
  {\bibfnamefont{K.}~\bibnamefont{Hicks}}, \bibinfo {author}
  {\bibfnamefont{E.~R.}\ \bibnamefont{Miller}}, \bibinfo {author}
  {\bibfnamefont{S.}~\bibnamefont{Persey}}, \bibinfo {author}
  {\bibfnamefont{O.}~\bibnamefont{Alinvi}},\ and\ \bibinfo {author}
  {\bibfnamefont{D.~J.~T.}\ \bibnamefont{Sumpter}},\ }%
  \bibfield{journal}{%
  \Doi{10.1007/s00265-008-0680-7}{\bibinfo {journal} {Behavioral Ecology and
  Sociobiology}}\ }%
  \textbf{\bibinfo {volume} {63}},\ \bibinfo {pages} {451} (\bibinfo {year}
  {2009})%
  \bibAnnoteFile{NoStop}{buhl009}%
\bibitem{perna012}%
  \BibitemOpen
  \bibfield{author}{%
  \bibinfo {author} {\bibfnamefont{A.}~\bibnamefont{Perna}}, \bibinfo {author}
  {\bibfnamefont{B.}~\bibnamefont{Granovskiy}}, \bibinfo {author}
  {\bibfnamefont{S.}~\bibnamefont{Garnier}}, \bibinfo {author}
  {\bibfnamefont{S.~C.}\ \bibnamefont{Nicolis}}, \bibinfo {author}
  {\bibfnamefont{M.}~\bibnamefont{Labédan}}, \bibinfo {author}
  {\bibfnamefont{G.}~\bibnamefont{Theraulaz}}, \bibinfo {author}
  {\bibfnamefont{V.}~\bibnamefont{Fourcassié}},\ and\ \bibinfo {author}
  {\bibfnamefont{D.~J.~T.}\ \bibnamefont{Sumpter}},\ }%
  \bibfield{journal}{%
  \Doi{10.1371/journal.pcbi.1002592}{\bibinfo {journal} {PLoS Computational
  Biology}}\ }%
  \textbf{\bibinfo {volume} {8}},\ \bibinfo {pages} {e1002592} (\bibinfo {year}
  {2012})%
  \bibAnnoteFile{NoStop}{perna012}%
\bibitem{griffon015}%
  \BibitemOpen
  \bibfield{author}{%
  \bibinfo {author} {\bibfnamefont{D.}~\bibnamefont{Griffon}}, \bibinfo
  {author} {\bibfnamefont{C.}~\bibnamefont{Andara}},\ and\ \bibinfo {author}
  {\bibfnamefont{K.}~\bibnamefont{Jaffe}},\ }%
  \emph{\bibinfo {title} {Emergence, self-organization and network efficiency
  in gigantic termite-nest-networks build using simple rules}},\ \bibinfo
  {type} {Tech. Rep.}\ \bibinfo {number} {arXiv:1506.01487}\ (\bibinfo
  {institution} {arXiv},\ \bibinfo {year} {2015})\
  \url{http://arxiv.org/pdf/1506.01487v1.pdf}%
  \bibAnnoteFile{NoStop}{griffon015}%
\bibitem{buhl006}%
  \BibitemOpen
  \bibfield{author}{%
  \bibinfo {author} {\bibfnamefont{J.}~\bibnamefont{Buhl}}, \bibinfo {author}
  {\bibfnamefont{J.}~\bibnamefont{Gautrais}}, \bibinfo {author}
  {\bibfnamefont{J.}~\bibnamefont{Louis~Deneubourg}}, \bibinfo {author}
  {\bibfnamefont{P.}~\bibnamefont{Kuntz}},\ and\ \bibinfo {author}
  {\bibfnamefont{G.}~\bibnamefont{Theraulaz}},\ }%
  \bibfield{journal}{%
  \Doi{10.1016/j.jtbi.2006.06.018}{\bibinfo {journal} {J Theor Biol}}\ }%
  \textbf{\bibinfo {volume} {243}},\ \bibinfo {pages} {287} (\bibinfo {month}
  {Dec}\ \bibinfo {year} {2006})%
  \bibAnnoteFile{NoStop}{buhl006}%
\bibitem{lee006}%
  \BibitemOpen
  \bibfield{author}{%
  \bibinfo {author} {\bibfnamefont{S.-H.}\ \bibnamefont{Lee}}, \bibinfo
  {author} {\bibfnamefont{P.}~\bibnamefont{Bardunias}},\ and\ \bibinfo {author}
  {\bibfnamefont{N.-Y.}\ \bibnamefont{Su}},\ }%
  \bibfield{journal}{%
  \Doi{10.1016/j.jtbi.2006.07.026}{\bibinfo {journal} {Journal of Theoretical
  Biology}}\ }%
  \textbf{\bibinfo {volume} {243}},\ \bibinfo {pages} {493} (\bibinfo {year}
  {2006})%
  \bibAnnoteFile{NoStop}{lee006}%
\bibitem{haifig011}%
  \BibitemOpen
  \bibfield{author}{%
  \bibinfo {author} {\bibfnamefont{I.}~\bibnamefont{Haifig}}, \bibinfo {author}
  {\bibfnamefont{C.}~\bibnamefont{Jost}}, \bibinfo {author}
  {\bibfnamefont{V.}~\bibnamefont{Janei}},\ and\ \bibinfo {author}
  {\bibfnamefont{A.~M.}\ \bibnamefont{Costa-Leonardo}},\ }%
  \bibfield{journal}{%
  \Doi{10.1016/j.anbehav.2011.09.025}{\bibinfo {journal} {Animal Behaviour}}\
  }%
  \textbf{\bibinfo {volume} {82}},\ \bibinfo {pages} {1409} (\bibinfo {year}
  {2011})%
  \bibAnnoteFile{NoStop}{haifig011}%
\bibitem{jost012}%
  \BibitemOpen
  \bibfield{author}{%
  \bibinfo {author} {\bibfnamefont{C.}~\bibnamefont{Jost}}, \bibinfo {author}
  {\bibfnamefont{I.}~\bibnamefont{Haifig}}, \bibinfo {author}
  {\bibfnamefont{C.~R.~R.}\ \bibnamefont{de~Camargo-Dietrich}},\ and\ \bibinfo
  {author} {\bibfnamefont{A.~M.}\ \bibnamefont{Costa-Leonardo}},\ }%
  \bibfield{journal}{%
  \Doi{10.1007/s00040-012-0229-7}{\bibinfo {journal} {Insectes Sociaux}}\ }%
  \textbf{\bibinfo {volume} {59}},\ \bibinfo {pages} {369} (\bibinfo {year}
  {2012})%
  \bibAnnoteFile{NoStop}{jost012}%
\bibitem{camazine001}%
  \BibitemOpen
  \bibfield{author}{%
  \bibinfo {author} {\bibfnamefont{S.}~\bibnamefont{Camazine}}, \bibinfo
  {author} {\bibfnamefont{J.-L.}\ \bibnamefont{Deneubourg}}, \bibinfo {author}
  {\bibfnamefont{N.~R.}\ \bibnamefont{Franks}}, \bibinfo {author}
  {\bibfnamefont{J.}~\bibnamefont{Sneyd}}, \bibinfo {author}
  {\bibfnamefont{G.}~\bibnamefont{Theraulaz}},\ and\ \bibinfo {author}
  {\bibfnamefont{E.}~\bibnamefont{Bonabeau}},\ }%
  \emph{\bibinfo {title} {Self-organization in biological systems}}\ (\bibinfo
  {publisher} {Princeton University Press},\ \bibinfo {address} {Princeton},\
  \bibinfo {year} {2001})\ p.\ \bibinfo {pages} {538}%
  \bibAnnoteFile{NoStop}{camazine001}%
\bibitem{levinson006}%
  \BibitemOpen
  \bibfield{author}{%
  \bibinfo {author} {\bibfnamefont{D.}~\bibnamefont{Levinson}}\ and\ \bibinfo
  {author} {\bibfnamefont{B.}~\bibnamefont{Yerra}},\ }%
  \bibfield{journal}{%
  \Doi{10.1287/trsc.1050.0132}{\bibinfo {journal} {Transportation Science}}\ }%
  \textbf{\bibinfo {volume} {40}},\ \bibinfo {pages} {179} (\bibinfo {year}
  {2006})%
  \bibAnnoteFile{NoStop}{levinson006}%
\bibitem{franks91}%
  \BibitemOpen
  \bibfield{author}{%
  \bibinfo {author} {\bibfnamefont{N.~R.}\ \bibnamefont{Franks}}, \bibinfo
  {author} {\bibfnamefont{N.}~\bibnamefont{Gomez}}, \bibinfo {author}
  {\bibfnamefont{S.}~\bibnamefont{Goss}},\ and\ \bibinfo {author}
  {\bibfnamefont{J.-L.}\ \bibnamefont{Deneubourg}},\ }%
  \bibfield{journal}{%
  \bibinfo {journal} {Journal of Insect Behavior}\ }%
  \textbf{\bibinfo {volume} {4}},\ \bibinfo {pages} {583} (\bibinfo {year}
  {1991})%
  \bibAnnoteFile{NoStop}{franks91}%
\bibitem{gautrais014}%
  \BibitemOpen
  \bibfield{author}{%
  \bibinfo {author} {\bibfnamefont{J.}~\bibnamefont{Gautrais}}, \bibinfo
  {author} {\bibfnamefont{J.}~\bibnamefont{Buhl}}, \bibinfo {author}
  {\bibfnamefont{S.}~\bibnamefont{Valverde}}, \bibinfo {author}
  {\bibfnamefont{P.}~\bibnamefont{Kuntz}},\ and\ \bibinfo {author}
  {\bibfnamefont{G.}~\bibnamefont{Theraulaz}},\ }%
  \bibfield{journal}{%
  \Doi{10.1371/journal.pone.0109436}{\bibinfo {journal} {PLoS One}}\ }%
  \textbf{\bibinfo {volume} {9}},\ \bibinfo {pages} {e109436} (\bibinfo {year}
  {2014})%
  \bibAnnoteFile{NoStop}{gautrais014}%
\bibitem{strano2012elementary}%
  \BibitemOpen
  \bibfield{author}{%
  \bibinfo {author} {\bibfnamefont{E.}~\bibnamefont{Strano}}, \bibinfo {author}
  {\bibfnamefont{V.}~\bibnamefont{Nicosia}}, \bibinfo {author}
  {\bibfnamefont{V.}~\bibnamefont{Latora}}, \bibinfo {author}
  {\bibfnamefont{S.}~\bibnamefont{Porta}},\ and\ \bibinfo {author}
  {\bibfnamefont{M.}~\bibnamefont{Barthélemy}},\ }%
  \bibfield{journal}{%
  \bibinfo {journal} {Scientific reports}\ }%
  \textbf{\bibinfo {volume} {2}} (\bibinfo {year} {2012})%
  \bibAnnoteFile{NoStop}{strano2012elementary}%
\bibitem{tero006}%
  \BibitemOpen
  \bibfield{author}{%
  \bibinfo {author} {\bibfnamefont{A.}~\bibnamefont{Tero}}, \bibinfo {author}
  {\bibfnamefont{R.}~\bibnamefont{Kobayashi}},\ and\ \bibinfo {author}
  {\bibfnamefont{T.}~\bibnamefont{Nakagaki}},\ }%
  \bibfield{booktitle}{%
  \emph{\bibinfo {booktitle} {Information and Material Flows in Complex
  Networks Information and Material Flows in Complex Networks}},\ }%
  \bibfield{journal}{%
  \Doi{http://dx.doi.org/10.1016/j.physa.2006.01.053}{\bibinfo {journal}
  {Physica A: Statistical Mechanics and its Applications}}\ }%
  \textbf{\bibinfo {volume} {363}},\ \bibinfo {pages} {115} (\bibinfo {month}
  {4}\ \bibinfo {year} {2006}),\
  \url{http://www.sciencedirect.com/science/article/pii/S0378437106000963}%
  \bibAnnoteFile{NoStop}{tero006}%
\bibitem{ma013}%
  \BibitemOpen
  \bibfield{author}{%
  \bibinfo {author} {\bibfnamefont{Q.}~\bibnamefont{Ma}}, \bibinfo {author}
  {\bibfnamefont{A.}~\bibnamefont{Johansson}}, \bibinfo {author}
  {\bibfnamefont{A.}~\bibnamefont{Tero}}, \bibinfo {author}
  {\bibfnamefont{T.}~\bibnamefont{Nakagaki}},\ and\ \bibinfo {author}
  {\bibfnamefont{D.~J.~T.}\ \bibnamefont{Sumpter}},\ }%
  \bibfield{journal}{%
  \Doi{10.1098/rsif.2012.0864}{\bibinfo {journal} {Journal of the Royal Society
  Interface}}\ }%
  \textbf{\bibinfo {volume} {10}},\ \bibinfo {pages} {20120864} (\bibinfo
  {year} {2013})%
  \bibAnnoteFile{NoStop}{ma013}%
\bibitem{valverde009}%
  \BibitemOpen
  \bibfield{author}{%
  \bibinfo {author} {\bibfnamefont{S.}~\bibnamefont{Valverde}}, \bibinfo
  {author} {\bibfnamefont{B.}~\bibnamefont{Corominas-Murtra}}, \bibinfo
  {author} {\bibfnamefont{A.}~\bibnamefont{Perna}}, \bibinfo {author}
  {\bibfnamefont{P.}~\bibnamefont{Kuntz}}, \bibinfo {author}
  {\bibfnamefont{G.}~\bibnamefont{Theraulaz}},\ and\ \bibinfo {author}
  {\bibfnamefont{R.~V.}\ \bibnamefont{Sole}},\ }%
  \bibfield{journal}{%
  \Doi{10.1103/PhysRevE.79.066106}{\bibinfo {journal} {Physical Review E}}\ }%
  \textbf{\bibinfo {volume} {79}},\ \bibinfo {pages} {066106} (\bibinfo {year}
  {2009})%
  \bibAnnoteFile{NoStop}{valverde009}%
\bibitem{barthelemy006}%
  \BibitemOpen
  \bibfield{author}{%
  \bibinfo {author} {\bibfnamefont{M.}~\bibnamefont{Barth\'elemy}}\ and\
  \bibinfo {author} {\bibfnamefont{A.}~\bibnamefont{Flammini}},\ }%
  \bibfield{journal}{%
  \Doi{10.1088/1742-5468/2006/07/L07002}{\bibinfo {journal} {Journal of
  Statistical Mechanics: Theory and Experiment}}\ }%
  \textbf{\bibinfo {volume} {2006}},\ \bibinfo {pages} {L07002} (\bibinfo
  {year} {2006}),\ \url{http://stacks.iop.org/1742-5468/2006/i=07/a=L07002}%
  \bibAnnoteFile{NoStop}{barthelemy006}%
\bibitem{barthelemy008}%
  \BibitemOpen
  \bibfield{author}{%
  \bibinfo {author} {\bibfnamefont{M.}~\bibnamefont{Barth\'elemy}}\ and\
  \bibinfo {author} {\bibfnamefont{A.}~\bibnamefont{Flammini}},\ }%
  \bibfield{journal}{%
  \Doi{10.1103/PhysRevLett.100.138702}{\bibinfo {journal} {Physical Review
  Letters}}\ }%
  \textbf{\bibinfo {volume} {100}},\ \bibinfo {pages} {138702} (\bibinfo {year}
  {2008})%
  \bibAnnoteFile{NoStop}{barthelemy008}%
\bibitem{Itzkovitz2005}%
  \BibitemOpen
  \bibfield{author}{%
  \bibinfo {author} {\bibfnamefont{S.}~\bibnamefont{Itzkovitz}}\ and\ \bibinfo
  {author} {\bibfnamefont{U.}~\bibnamefont{Alon}},\ }%
  \bibfield{journal}{%
  \Doi{10.1103/PhysRevE.71.026117}{\bibinfo {journal} {Physical Review E}}\ }%
  \textbf{\bibinfo {volume} {71}},\ \bibinfo {pages} {026117} (\bibinfo {month}
  {Part 2 FEB}\ \bibinfo {year} {2005}),\
  \url{http://dx.doi.org/10.1103/PhysRevE.71.026117}%
  \bibAnnoteFile{NoStop}{Itzkovitz2005}%
\bibitem{albert002}%
  \BibitemOpen
  \bibfield{author}{%
  \bibinfo {author} {\bibfnamefont{R.}~\bibnamefont{Albert}}\ and\ \bibinfo
  {author} {\bibfnamefont{A.-L.}\ \bibnamefont{Barabási}},\ }%
  \bibfield{journal}{%
  \bibinfo {journal} {Reviews of modern physics}\ }%
  \textbf{\bibinfo {volume} {74}},\ \bibinfo {pages} {47} (\bibinfo {year}
  {2002})%
  \bibAnnoteFile{NoStop}{albert002}%
\bibitem{newman003pub}%
  \BibitemOpen
  \bibfield{author}{%
  \bibinfo {author} {\bibfnamefont{M.~E.~J.}\ \bibnamefont{Newman}},\ }%
  \bibfield{journal}{%
  \Doi{10.1137/S003614450342480}{\bibinfo {journal} {SIAM Rev}}\ }%
  \textbf{\bibinfo {volume} {45}},\ \bibinfo {pages} {167} (\bibinfo {year}
  {2003}),\ \url{http://arxiv.org/pdf/cond-mat/0303516.pdf}%
  \bibAnnoteFile{NoStop}{newman003pub}%
\bibitem{costa011}%
  \BibitemOpen
  \bibfield{author}{%
  \bibinfo {author} {\bibfnamefont{L.~d.~F.}\ \bibnamefont{Costa}}, \bibinfo
  {author} {\bibfnamefont{O.~N.}\ \bibnamefont{Oliveira~Jr}},\ and\ \bibinfo
  {author} {\bibfnamefont{G.}~\bibnamefont{Travieso}},\ }%
  \bibfield{journal}{%
  \Doi{10.1080/00018732.2011.572452}{\bibinfo {journal} {Advances in Physics}}\
  }%
  \textbf{\bibinfo {volume} {60}},\ \bibinfo {pages} {329} (\bibinfo {year}
  {2011})%
  \bibAnnoteFile{NoStop}{costa011}%
\bibitem{barthelemy010pub}%
  \BibitemOpen
  \bibfield{author}{%
  \bibinfo {author} {\bibfnamefont{M.}~\bibnamefont{Barth\'elemy}},\ }%
  \bibfield{journal}{%
  \Doi{10.1016/j.physrep.2010.11.002}{\bibinfo {journal} {Physics Reports}}\ }%
  \textbf{\bibinfo {volume} {499}},\ \bibinfo {pages} {1} (\bibinfo {year}
  {2011}),\ \url{http://arxiv.org/abs/1010.0302v2}%
  \bibAnnoteFile{NoStop}{barthelemy010pub}%
\bibitem{xie009}%
  \BibitemOpen
  \bibfield{author}{%
  \bibinfo {author} {\bibfnamefont{F.}~\bibnamefont{Xie}}\ and\ \bibinfo
  {author} {\bibfnamefont{D.}~\bibnamefont{Levinson}},\ }%
  \bibfield{journal}{%
  \Doi{10.1007/s11067-007-9037-4}{\bibinfo {journal} {Modeling the Growth of
  Transportation Networks: A Comprehensive Review Netw Spat Econ}}\ }%
  \textbf{\bibinfo {volume} {9}},\ \bibinfo {pages} {291} (\bibinfo {year}
  {2009})%
  \bibAnnoteFile{NoStop}{xie009}%
\bibitem{jmhasly99}%
  \BibitemOpen
  \bibfield{author}{%
  \bibinfo {author} {\bibfnamefont{P.}~\bibnamefont{Jmhasly}}\ and\ \bibinfo
  {author} {\bibfnamefont{R.~H.}\ \bibnamefont{Leuthold}},\ }%
  \bibfield{journal}{%
  \Doi{10.1007/s000400050154}{\bibinfo {journal} {Insectes Sociaux}}\ }%
  \textbf{\bibinfo {volume} {46}},\ \bibinfo {pages} {332} (\bibinfo {year}
  {1999})%
  \bibAnnoteFile{NoStop}{jmhasly99}%
\bibitem{noirot00}%
  \BibitemOpen
  \bibfield{author}{%
  \bibinfo {author} {\bibfnamefont{C.}~\bibnamefont{Noirot}}\ and\ \bibinfo
  {author} {\bibfnamefont{J.}~\bibnamefont{Darlington}},\ }%
  in\ \emph{\bibinfo {booktitle} {Termites: evolution, sociality, symbioses,
  ecology}},\ \bibinfo {editor} {edited by\ \bibinfo {editor}
  {\bibfnamefont{T.}~\bibnamefont{Abe}}, \bibinfo {editor}
  {\bibfnamefont{D.~E.}\ \bibnamefont{Bignell}},\ and\ \bibinfo {editor}
  {\bibfnamefont{M.}~\bibnamefont{Higashi}}}\ (\bibinfo {publisher} {Kluwer
  Academic Publishers Dordrecht},\ \bibinfo {year} {2000})\ Chap.~\bibinfo
  {chapter} {6}, pp.\ \bibinfo {pages} {121--139}%
  \bibAnnoteFile{NoStop}{noirot00}%
\bibitem{tschinkel010b}%
  \BibitemOpen
  \bibfield{author}{%
  \bibinfo {author} {\bibfnamefont{W.~R.}\ \bibnamefont{Tschinkel}},\ }%
  \bibfield{journal}{%
  \Doi{10.1673/031.010.6501}{\bibinfo {journal} {Journal of Insect Science}}\
  }%
  \textbf{\bibinfo {volume} {10}},\ \bibinfo {pages} {1} (\bibinfo {year}
  {2010})%
  \bibAnnoteFile{NoStop}{tschinkel010b}%
\bibitem{perna008a}%
  \BibitemOpen
  \bibfield{author}{%
  \bibinfo {author} {\bibfnamefont{A.}~\bibnamefont{Perna}}, \bibinfo {author}
  {\bibfnamefont{C.}~\bibnamefont{Jost}}, \bibinfo {author}
  {\bibfnamefont{E.}~\bibnamefont{Couturier}}, \bibinfo {author}
  {\bibfnamefont{S.}~\bibnamefont{Valverde}}, \bibinfo {author}
  {\bibfnamefont{S.}~\bibnamefont{Douady}},\ and\ \bibinfo {author}
  {\bibfnamefont{G.}~\bibnamefont{Theraulaz}},\ }%
  \bibfield{journal}{%
  \Doi{10.1007/s00114-008-0388-6}{\bibinfo {journal} {Naturwissenschaften}}\ }%
  \textbf{\bibinfo {volume} {95}},\ \bibinfo {pages} {877} (\bibinfo {year}
  {2008})%
  \bibAnnoteFile{NoStop}{perna008a}%
\bibitem{SM}%
  \BibitemOpen
  \enquote{\bibinfo {title} {See supplemental material at
  \url{http://link.aps.org/supplemental/ 10.1103/PhysRevE.92.062810} for a
  detailed statistical description of each nest},}\ %
  \bibAnnoteFile{NoStop}{SM}%
\bibitem{noirot70}%
  \BibitemOpen
  \bibfield{author}{%
  \bibinfo {author} {\bibfnamefont{C.}~\bibnamefont{Noirot}},\ }%
  in\ \emph{\bibinfo {booktitle} {Biology of Termites vol II}},\ \bibinfo
  {editor} {edited by\ \bibinfo {editor}
  {\bibfnamefont{K.}~\bibnamefont{Krishna}}\ and\ \bibinfo {editor}
  {\bibfnamefont{F.~M.}\ \bibnamefont{Weesner}}}\ (\bibinfo {publisher}
  {Academic Press},\ \bibinfo {year} {1970})\ pp.\ \bibinfo {pages} {73--125}%
  \bibAnnoteFile{NoStop}{noirot70}%
\bibitem{josens010}%
  \BibitemOpen
  \bibfield{author}{%
  \bibinfo {author} {\bibfnamefont{G.}~\bibnamefont{Josens}}\ and\ \bibinfo
  {author} {\bibfnamefont{K.}~\bibnamefont{Soki}},\ }%
  \bibfield{journal}{%
  \Doi{10.1007/s00040-010-0085-2}{\bibinfo {journal} {Insectes Sociaux}}\ }%
  \textbf{\bibinfo {volume} {57}},\ \bibinfo {pages} {303} (\bibinfo {year}
  {2010})%
  \bibAnnoteFile{NoStop}{josens010}%
\bibitem{bruinsma79}%
  \BibitemOpen
  \bibfield{author}{%
  \bibinfo {author} {\bibfnamefont{O.~H.}\ \bibnamefont{Bruinsma}},\ }%
  \emph{\bibinfo {title} {An analysis of building behaviour of the termite
  \emph{Macrotermes subhyalinus}}},\ Ph.D. thesis,\ \bibinfo {school}
  {Lanbouwhoogeschool te Wageningen}, \bibinfo {address} {Belgium} (\bibinfo
  {year} {1979})%
  \bibAnnoteFile{NoStop}{bruinsma79}%
\bibitem{jander74}%
  \BibitemOpen
  \bibfield{author}{%
  \bibinfo {author} {\bibfnamefont{R.}~\bibnamefont{Jander}}\ and\ \bibinfo
  {author} {\bibfnamefont{K.}~\bibnamefont{Daumer}},\ }%
  \bibfield{journal}{%
  \bibinfo {journal} {Insectes Sociaux}\ }%
  \textbf{\bibinfo {volume} {21}},\ \bibinfo {pages} {45} (\bibinfo {year}
  {1974})%
  \bibAnnoteFile{NoStop}{jander74}%
\bibitem{girvannewman002}%
  \BibitemOpen
  \bibfield{author}{%
  \bibinfo {author} {\bibfnamefont{M.}~\bibnamefont{Girvan}}\ and\ \bibinfo
  {author} {\bibfnamefont{M.~E.~J.}\ \bibnamefont{Newman}},\ }%
  \bibfield{journal}{%
  \Doi{10.1073/pnas.122653799}{\bibinfo {journal} {Proc Natl Acad Sci U S A}}\
  }%
  \textbf{\bibinfo {volume} {99}},\ \bibinfo {pages} {7821} (\bibinfo {month}
  {Jun}\ \bibinfo {year} {2002})%
  \bibAnnoteFile{NoStop}{girvannewman002}%
\bibitem{bardunias009a}%
  \BibitemOpen
  \bibfield{author}{%
  \bibinfo {author} {\bibfnamefont{P.}~\bibnamefont{Bardunias}}\ and\ \bibinfo
  {author} {\bibfnamefont{N.-Y.}\ \bibnamefont{Su}},\ }%
  \bibfield{journal}{%
  \Doi{10.1016/j.anbehav.2009.06.024}{\bibinfo {journal} {Animal Behaviour}}\
  }%
  \textbf{\bibinfo {volume} {78}},\ \bibinfo {pages} {755} (\bibinfo {year}
  {2009})%
  \bibAnnoteFile{NoStop}{bardunias009a}%
\bibitem{perna008b}%
  \BibitemOpen
  \bibfield{author}{%
  \bibinfo {author} {\bibfnamefont{A.}~\bibnamefont{Perna}}, \bibinfo {author}
  {\bibfnamefont{S.}~\bibnamefont{Valverde}}, \bibinfo {author}
  {\bibfnamefont{J.}~\bibnamefont{Gautrais}}, \bibinfo {author}
  {\bibfnamefont{C.}~\bibnamefont{Jost}}, \bibinfo {author}
  {\bibfnamefont{R.~V.}\ \bibnamefont{Solé}}, \bibinfo {author}
  {\bibfnamefont{P.}~\bibnamefont{Kuntz}},\ and\ \bibinfo {author}
  {\bibfnamefont{G.}~\bibnamefont{Theraulaz}},\ }%
  \bibfield{journal}{%
  \Doi{10.1016/j.physa.2008.07/019}{\bibinfo {journal} {Physica A}}\ }%
  \textbf{\bibinfo {volume} {387}},\ \bibinfo {pages} {6235} (\bibinfo {year}
  {2008})%
  \bibAnnoteFile{NoStop}{perna008b}%
\bibitem{oppenheim1991cell}%
  \BibitemOpen
  \bibfield{author}{%
  \bibinfo {author} {\bibfnamefont{R.~W.}\ \bibnamefont{Oppenheim}},\ }%
  \bibfield{journal}{%
  \bibinfo {journal} {Annual review of neuroscience}\ }%
  \textbf{\bibinfo {volume} {14}},\ \bibinfo {pages} {453} (\bibinfo {year}
  {1991})%
  \bibAnnoteFile{NoStop}{oppenheim1991cell}%
\bibitem{Chechik1998}%
  \BibitemOpen
  \bibfield{author}{%
  \bibinfo {author} {\bibfnamefont{G.}~\bibnamefont{Chechik}}, \bibinfo
  {author} {\bibfnamefont{I.}~\bibnamefont{Meilijson}},\ and\ \bibinfo {author}
  {\bibfnamefont{E.}~\bibnamefont{Ruppin}},\ }%
  \bibfield{booktitle}{%
  \emph{\bibinfo {booktitle} {Neural Computation}},\ }%
  \bibfield{journal}{%
  \Doi{10.1162/089976698300017124}{\bibinfo {journal} {Neural Computation}}\ }%
  \textbf{\bibinfo {volume} {10}},\ \bibinfo {pages} {1759} (\bibinfo {month}
  {Oct.}\ \bibinfo {year} {1998}),\ ISSN \bibinfo {issn} {0899-7667},\
  \url{http://dx.doi.org/10.1162/089976698300017124}%
  \bibAnnoteFile{NoStop}{Chechik1998}%
\bibitem{Lew2011}%
  \BibitemOpen
  \bibfield{author}{%
  \bibinfo {author} {\bibfnamefont{R.~R.}\ \bibnamefont{Lew}},\ }%
  \bibfield{journal}{%
  \bibinfo {journal} {Nat Rev Micro}\ }%
  \textbf{\bibinfo {volume} {9}},\ \bibinfo {pages} {509} (\bibinfo {month}
  {Jul.}\ \bibinfo {year} {2011}),\ ISSN \bibinfo {issn} {1740-1526},\
  \url{http://dx.doi.org/10.1038/nrmicro2591}%
  \bibAnnoteFile{NoStop}{Lew2011}%
\bibitem{falconer2005biomass}%
  \BibitemOpen
  \bibfield{author}{%
  \bibinfo {author} {\bibfnamefont{R.~E.}\ \bibnamefont{Falconer}}, \bibinfo
  {author} {\bibfnamefont{J.~L.}\ \bibnamefont{Bown}}, \bibinfo {author}
  {\bibfnamefont{N.~A.}\ \bibnamefont{White}},\ and\ \bibinfo {author}
  {\bibfnamefont{J.~W.}\ \bibnamefont{Crawford}},\ }%
  \bibfield{journal}{%
  \bibinfo {journal} {Proceedings of the Royal Society B: Biological Sciences}\
  }%
  \textbf{\bibinfo {volume} {272}},\ \bibinfo {pages} {1727} (\bibinfo {year}
  {2005})%
  \bibAnnoteFile{NoStop}{falconer2005biomass}%
\bibitem{kessler82}%
  \BibitemOpen
  \bibfield{author}{%
  \bibinfo {author} {\bibfnamefont{D.}~\bibnamefont{Kessler}},\ }%
  in\ \emph{\bibinfo {booktitle} {Cell Biology of \emph{Physarum} and
  \emph{Didymium}}},\ Vol.~\bibinfo {volume} {1},\ \bibinfo {editor} {edited
  by\ \bibinfo {editor} {\bibfnamefont{C.}~\bibnamefont{Aldrich}}\ and\
  \bibinfo {editor} {\bibfnamefont{J.~W.}\ \bibnamefont{Daniel}}}\ (\bibinfo
  {publisher} {Academic Press},\ \bibinfo {address} {New York},\ \bibinfo
  {year} {1982})\ pp.\ \bibinfo {pages} {145--208}%
  \bibAnnoteFile{NoStop}{kessler82}%
\bibitem{toffin009}%
  \BibitemOpen
  \bibfield{author}{%
  \bibinfo {author} {\bibfnamefont{E.}~\bibnamefont{Toffin}}, \bibinfo {author}
  {\bibfnamefont{D.}~\bibnamefont{Di~Paolo}}, \bibinfo {author}
  {\bibfnamefont{A.}~\bibnamefont{Campo}}, \bibinfo {author}
  {\bibfnamefont{C.}~\bibnamefont{Detrain}},\ and\ \bibinfo {author}
  {\bibfnamefont{J.-L.}\ \bibnamefont{Deneubourg}},\ }%
  \bibfield{journal}{%
  \Doi{10.1073/pnas.0902685106}{\bibinfo {journal} {Proc Natl Acad Sci U S A}}\
  }%
  \textbf{\bibinfo {volume} {106}},\ \bibinfo {pages} {18616} (\bibinfo {month}
  {Nov}\ \bibinfo {year} {2009})%
  \bibAnnoteFile{NoStop}{toffin009}%
\bibitem{tschinkel010}%
  \BibitemOpen
  \bibfield{author}{%
  \bibinfo {author} {\bibfnamefont{W.~R.}\ \bibnamefont{Tschinkel}},\ }%
  \bibfield{journal}{%
  \Doi{10.1673/031.010.8801}{\bibinfo {journal} {J Insect Sci}}\ }%
  \textbf{\bibinfo {volume} {10}},\ \bibinfo {pages} {88} (\bibinfo {year}
  {2010})%
  \bibAnnoteFile{NoStop}{tschinkel010}%
\bibitem{tschinkel011}%
  \BibitemOpen
  \bibfield{author}{%
  \bibinfo {author} {\bibfnamefont{W.~R.}\ \bibnamefont{Tschinkel}},\ }%
  \bibfield{journal}{%
  \Doi{10.1673/031.011.10501}{\bibinfo {journal} {J Insect Sci}}\ }%
  \textbf{\bibinfo {volume} {11}},\ \bibinfo {pages} {105} (\bibinfo {year}
  {2011})%
  \bibAnnoteFile{NoStop}{tschinkel011}%
\bibitem{minter012}%
  \BibitemOpen
  \bibfield{author}{%
  \bibinfo {author} {\bibfnamefont{N.~J.}\ \bibnamefont{Minter}}, \bibinfo
  {author} {\bibfnamefont{N.~R.}\ \bibnamefont{Franks}},\ and\ \bibinfo
  {author} {\bibfnamefont{K.~A.~R.}\ \bibnamefont{Brown}},\ }%
  \bibfield{journal}{%
  \Doi{10.1098/rsif.2011.0377}{\bibinfo {journal} {J R Soc Interface}}\ }%
  \textbf{\bibinfo {volume} {9}},\ \bibinfo {pages} {586} (\bibinfo {month}
  {Mar}\ \bibinfo {year} {2012})%
  \bibAnnoteFile{NoStop}{minter012}%
\bibitem{moreira004}%
  \BibitemOpen
  \bibfield{author}{%
  \bibinfo {author} {\bibfnamefont{A.~A.}\ \bibnamefont{Moreira}}, \bibinfo
  {author} {\bibfnamefont{L.~C.}\ \bibnamefont{Forti}}, \bibinfo {author}
  {\bibfnamefont{A.~P.~P.}\ \bibnamefont{Andrade}}, \bibinfo {author}
  {\bibfnamefont{M.~A.~C.}\ \bibnamefont{Boaretto}},\ and\ \bibinfo {author}
  {\bibfnamefont{J.~F.~S.}\ \bibnamefont{Lopes}},\ }%
  \bibfield{journal}{%
  \bibinfo {journal} {Studies on Neotropical Fauna and Environment}\ }%
  \textbf{\bibinfo {volume} {39}},\ \bibinfo {pages} {109} (\bibinfo {year}
  {2004})%
  \bibAnnoteFile{NoStop}{moreira004}%
\end{thebibliography}%

\end{document}